\documentclass[
reprint,
amsmath,
amssymb,
aps,
nofootinbib,
]{revtex4-2}

\usepackage{graphicx}	
\usepackage{dcolumn}
\usepackage{bm}
\usepackage[hidelinks]{hyperref}
\usepackage{natbib}
\usepackage{xcolor}
\usepackage{booktabs}
\usepackage{placeins}

\begin{document}


\title{Bayesian constraints on the astrophysical neutrino source
  population \\ from IceCube data}

\author{F. Capel$^{1}$\email{E-mail: capel@kth.se (FC)},
  D. J. Mortlock$^{2,3,4}$ and C. Finley$^{5}$}

\affiliation{$^{1}$Department of Physics, KTH Royal Institute of
  Technology, and The Oskar Klein Centre, SE-10691 Stockholm, Sweden}
\affiliation{$^{2}$Astrophysics Group, Imperial College London,
  Blackett Laboratory, Prince Consort Road, London SW7 2AZ, UK}
\affiliation{$^{3}$Statistics Section, Department of Mathematics,
  Imperial College London, London SW7 2AZ, UK}
\affiliation{$^{4}$Department of Astronomy, Stockholm University,
  AlbaNova, SE-106 91 Stockholm, Sweden} \affiliation{$^{5}$The Oskar
  Klein Centre, Department of Physics, Stockholm University, AlbaNova,
  SE-10691 Stockholm, Sweden}

\date{\today}

\begin{abstract}
  \noindent
  We present constraints on an astrophysical population of neutrino
  sources imposed by recent data from the IceCube neutrino
  observatory. By using the IceCube point source search method to
  model the detection of sources, our detection criterion is more
  sensitive than using the observation of high-energy neutrino
  multiplets for source identification. We frame the problem as a
  Bayesian hierarchical model to connect the high-level population
  parameters to the IceCube data, allowing us to consistently account
  for all relevant sources of uncertainty in our model
  assumptions. Our results show that sources with a local density of
  $n_0 \gtrsim 10^{-7}$~$\mathrm{Mpc}^{-3}$ and luminosity
  $L \lesssim 10^{43}$~erg~$\mathrm{s}^{-1}$ are the most likely
  candidates, but that populations of rare sources with
  $n_0 \simeq 10^{-9}$~$\mathrm{Mpc}^{-3}$ and
  $L \simeq 10^{45}$~erg~$\mathrm{s}^{-1}$ can still be consistent
  with the IceCube observations. We demonstrate that these conclusions
  are strongly dependent on the source evolution considered, for which
  we consider a wide range of models. In doing so, we present
  realistic, model-independent constraints on the population
  parameters that reflect our current state of knowledge from
  astrophysical neutrino observations. We also use our framework to
  investigate constraints in the case of possible source detections
  and future instrument upgrades. Our approach is flexible and can be
  used to model specific source cases and extended to include
  multi-messenger information.
\end{abstract}

\maketitle

\section{\label{sec:intro}Introduction}

High energy neutrinos are expected to be produced through the hadronic
interactions of energetic cosmic rays with the matter and radiation
fields present in their astrophysical source environments. These
interactions lead to the production of charged and neutral pions,
which then decay into secondary gamma rays and neutrinos
\cite{Meszaros:2019gv, Halzen:2017ee}. Thanks to extensive
experimental efforts, it is now possible to detect high energy cosmic
rays and neutrinos in addition to electromagnetic radiation here on
Earth. These measurements provide complementary information that can
be harnessed to understand their common sources. Unlike charged
particles, neutrinos are weakly interacting and not deflected by
intervening magnetic fields, meaning that we can use neutrino
observations to identify cosmic-ray accelerators directly, as well as
to study distant sources and astrophysical processes occurring within
dense source environments.

A flux of high energy neutrinos ($>60$~TeV) has been measured by the
IceCube Collaboration \cite{Collaboration:2013hx,
  Aartsen:2015fr}. These astrophysical events have been identified by
their relatively high energies with respect to the atmospheric
background. This signal has now been detected with high significance
in both the ``high energy starting events''~\cite{Aartsen:2014fg} and
``through-going muons'' \cite{Aartsen:2016hp} channels. However, these
neutrinos seem to be isotropically distributed on the sky, and various
searches have not identified any point sources \cite{Abbasi:2011dl,
  Aartsen:2017fp, Collaboration:2019db, Aartsen:2019hi,
  Aartsen:2020fm}. Due to their isotropic nature, the majority of the
high energy IceCube neutrinos are thought to be of extra-Galactic
origin, with a Galactic contribution constrained to $\sim$~14\% above
1~TeV~\cite{Aartsen:2017by}.

Plausible extra-Galactic sources of the IceCube neutrinos are closely
connected to the sources of very high energy cosmic rays with energies
of up to 100~PeV, due to their production mechanism. Proposed sources
include different types of active galactic nuclei and galaxies with
high star formation rates as well as explosive transients such as
gamma-ray bursts and tidal disruption events
\cite{Meszaros:2017hm}. As neutrinos are weakly interacting and can
reach us from considerable distances, the bulk of the observed high
energy flux is expected to be due to the integrated contribution of
many distant sources. However, for certain source population
properties, we could be able to detect nearby, individual point
sources in addition to this diffuse component
\cite{Lipari:2006ck}. The combination of the measurement of an
astrophysical neutrino flux, but non-observation of individual
neutrino sources can be used to place constraints on the properties of
an unknown, steady-state source population \cite{Lipari:2008ew,
  Silvestri:2010ku, Murase:2012gy, Ahlers:2014em, Kowalski:2015gm,
  Murase:2016gly, Palladino:2019tm}. The general idea is that the
neutrino sources must collectively be able to satisfy the energy
density requirements of the observed astrophysical flux while being
consistent with the current non-detection of point sources. These
simple considerations can be used to test the validity of proposed
source models.

It is necessary to quantify source detection in order to place
constraints on the population from the non-observation of sources. In
previous work \cite{Lipari:2008ew, Ahlers:2014em, Kowalski:2015gm,
  Murase:2016gly, Palladino:2019tm}, this has been done by requiring a
point source to produce two or more high energy ($\sim 100$--200~TeV)
neutrinos in a detector. The argument is that at these energies, the
angular resolution is sufficiently high, and the atmospheric neutrino
background is sufficiently low, that the so-called neutrino
``multiplets,'' which have consistent arrival directions, must be
coming from a common source. However, the probability of high-energy
multiplets occurring by chance will become increasingly non-negligible
as the IceCube experiment continues to gather data, meaning that it
will be necessary to set even higher energy thresholds, consider only
multiplets including three or more events, or make stronger
corrections for the atmospheric neutrino backgrounds. All of these
approaches lead to broader constraints on the source population,
despite the increased detector exposure.

The goal of this work is to go beyond the multiplet assumption and
model the neutrino and source detection process to characterize the
detection probability in detail. We achieve this goal by connecting
the high-level population parameters to observable quantities through
the use of a Bayesian hierarchical model for the neutrino
sources. This approach allows us to bring together the IceCube
observations with a generic source population model in order to place
coherent constraints on the properties of the unknown sources. In this
way, the resulting constraints are summarized through the marginalized
posterior distribution over the relevant parameters. Our approach also
builds on previous work by including the significant effects of
uncertainties in the cosmological evolution of the unknown sources and
the IceCube observations, meaning that the final results reflect
realistic and model-independent assumptions. We present the physical
modeling assumptions in Section~\ref{sec:physical_model}, followed by
a description of neutrino detection with IceCube and the
characterization of the individual source detection probability in
Sections~\ref{sec:nu_detection_icecube} and
\ref{sec:detection_probability}. The statistical formalism for the
hierarchical model that we develop is introduced in
Section~\ref{sec:statistical_model}. We then apply our method to
recent IceCube observations in Section~\ref{sec:application} and
present the results in Section~\ref{sec:results}. Finally, we
summarize our framework and discuss some exciting possibilities for
its extension in Section~\ref{sec:conclusions}. The code used in this
work is open source and available
online\footnote{\url{www.github.com/cescalara/nu_pop} (to be made
  available upon publication).}.

\section{\label{sec:physical_model}Physical Model}

We consider a generic, extra-Galactic population of steady,
neutrino-emitting sources. Neutrinos are assumed to be produced by
individual point-like sources (Section~\ref{subsec:sources}), which
form a population with a density distribution that evolves over
cosmological scales (Section~\ref{subsec:population}). This population
will produce a flux of neutrinos that can be detected on Earth
(Section~\ref{subsec:flux_at_Earth}).

\subsection{\label{subsec:sources}Individual sources}

High energy neutrinos are expected to be produced through the
interactions of accelerated cosmic rays with their source
environments. As such, they should have a power-law energy spectrum
with a similar spectral index to that of the primary cosmic rays. We
assume that the neutrino spectrum is described by a power law, such
that the differential emission from a single neutrino source is given
by
\begin{equation}
  \frac{\mathrm{d}\bar{N}_\nu^\mathrm{src}}{\mathop{\mathrm{d}E} \mathop{\mathrm{d}t}} = L k_\gamma E_0^{-\gamma} \Bigg( \frac{E}{E_0} \Bigg)^{-\gamma},
  \label{eqn:point_source}
\end{equation}
where $\mathrm{d}\bar{N}_\nu^\mathrm{src} / \mathrm{d}E\mathrm{d}t$ is
the expected rate of neutrinos per unit energy, $\gamma$ is the
spectral index, and $E_0$ is the energy at which the power law is
normalized. The factor $k_\gamma$ is defined such that is the
luminosity between $E_\mathrm{min}$ and $E_\mathrm{max}$ is given by
\begin{equation}
  L = \int_{E_\mathrm{min}}^{E_\mathrm{max}} \mathop{\mathrm{d}E} E \frac{\mathrm{d}\bar{N}_\nu^\mathrm{src}}{\mathop{\mathrm{d}E} \mathop{\mathrm{d}t}}, 
  \label{eqn:luminosity}
\end{equation}
and so we have
\begin{equation}
  k_\gamma = \frac{(\gamma-2)E_\mathrm{min}^{\gamma-2}}{1 - (E_\mathrm{min}/E_\mathrm{max})^{\gamma-2}}.
\end{equation}
While this physical picture is clear for the case of sources that emit
neutrinos isotropically, this assumption is expected to be false for
many popular source candidate classes. For example, energetic blazars
could emit neutrinos that are strongly beamed along the axis of their
relativistic jets, such that all emission is contained within
$\Delta \Omega$. We handle this case by considering $L$ to be the
``isotropic-equivalent'' luminosity, such that
$L \equiv (4 \pi / \Delta \Omega)L_\mathrm{true}$ (see
e.g.,~\citet{Lipari:2006ck}). We assume that all sources produce
neutrinos independently of time according to their luminosity and
energy spectrum.

Here, we consider all neutrino sources to have the same
luminosity. While this is not realistic, this choice corresponds to
more conservative constraints on the source population from the
non-observation of point sources. The basic argument, as highlighted
in \citet{Bartos:2017es}, is that introducing a luminosity function
would increase the effective anisotropy of a population viewed from
Earth, therefore making it easier to detect individual neutrino
sources, and placing stronger constraints on the population for the
non-observation of point sources. Additionally, the neutrino
luminosity function is currently poorly constrained, given that the
details of the source physics and connection to observations at other
wavelengths remain open questions. However, it is potentially
interesting to consider a general, parametric luminosity function that
reflects our expectations. Such an extension of this model is
presented in Appendix~\ref{sec:lum_function}.

\subsection{\label{subsec:population}The source population}

We assume that individual sources are distributed isotropically
throughout the universe and that their density evolves over
cosmological scales. The structure of this cosmological evolution
depends on the category of source candidates considered, with popular
models including evolution following the star formation rate (SFR)
\citep{Madau:2014gh} or the AGN luminosity function
\cite{Aird:2010ko}. Flat or even negative evolutions are also often
considered for comparison, motivated by studies of the
ultra-high-energy cosmic ray data~\cite{Taylor:2015jx, Aab:2016zth,
  Batista:2019jg} and the debated evolution of certain types of BL Lac
objects~\citep{Ajello:2014fi}. In order to account for the uncertainty
in the exact form of the comoving density, we parameterize this as a
simple function capable of representing a range of relevant source
models, motivated by parameterizations used in \citet{Cole:2001uv} and
\citet{Ueda:kv}. The expected number of sources per unit comoving
volume is given by
\begin{equation}
  \frac{\mathrm{d}\bar{N}_\mathrm{s}^\mathrm{tot}}{\mathrm{d}V} = n_0 \frac{(1+z)^{p_1}}{(1 + z/z_c)^{p_2}} = n_0 f(z, \theta),
  \label{eqn:source_evolution}
\end{equation}
where $n_0$ is the local source density. The parameters $p_1$, $p_2$,
and $z_c$ can be varied to produce a wide range of possible source
evolutions, as shown in Figure~\ref{fig:source_evolution}. We refer to
these shape parameters collectively as $\theta = \{p_1, p_2, z_c\}$ in
the following sections for the sake of brevity, as they will typically
be treated as nuisance parameters and marginalized over. The expected
total number of sources in the universe can be found by integrating
the product of the density and comoving volume over redshift as
\begin{equation}
  \bar{N}_\mathrm{s}^\mathrm{tot} = \int_0^\infty \mathop{\mathrm{d}z} \frac{\mathrm{d}\bar{N}_\mathrm{s}^\mathrm{tot}}{\mathrm{d}V}\frac{\mathrm{d}V}{\mathrm{d}z}.
  \label{eqn:total_sources}
\end{equation}
The differential comoving volume is given by
\begin{equation}
  \frac{\mathrm{d}V}{\mathrm{d}z} = \frac{4\pi c D_L^2(z)}{H_0(1+z)^2\sqrt{\Omega_m(1+z)^3 + \Omega_\Lambda}},
\end{equation}
where $z$ is the redshift, $H_0$ is the Hubble constant, and $D_L$ is
the luminosity distance \cite{Peacock:2010aa}. We assume a flat
$\Lambda\mathrm{CDM}$ cosmology throughout this work, with
$H_0 = 70 \ \mathrm{km} \ \mathrm{s}^{-1} \ \mathrm{Mpc}^{-1}$,
$\Omega_m = 0.3$, and $\Omega_\Lambda = 0.7$.

\subsection{\label{subsec:flux_at_Earth}Incident neutrino flux at
  Earth}

\begin{figure}[h!]
  \centering \includegraphics[width=
  \columnwidth]{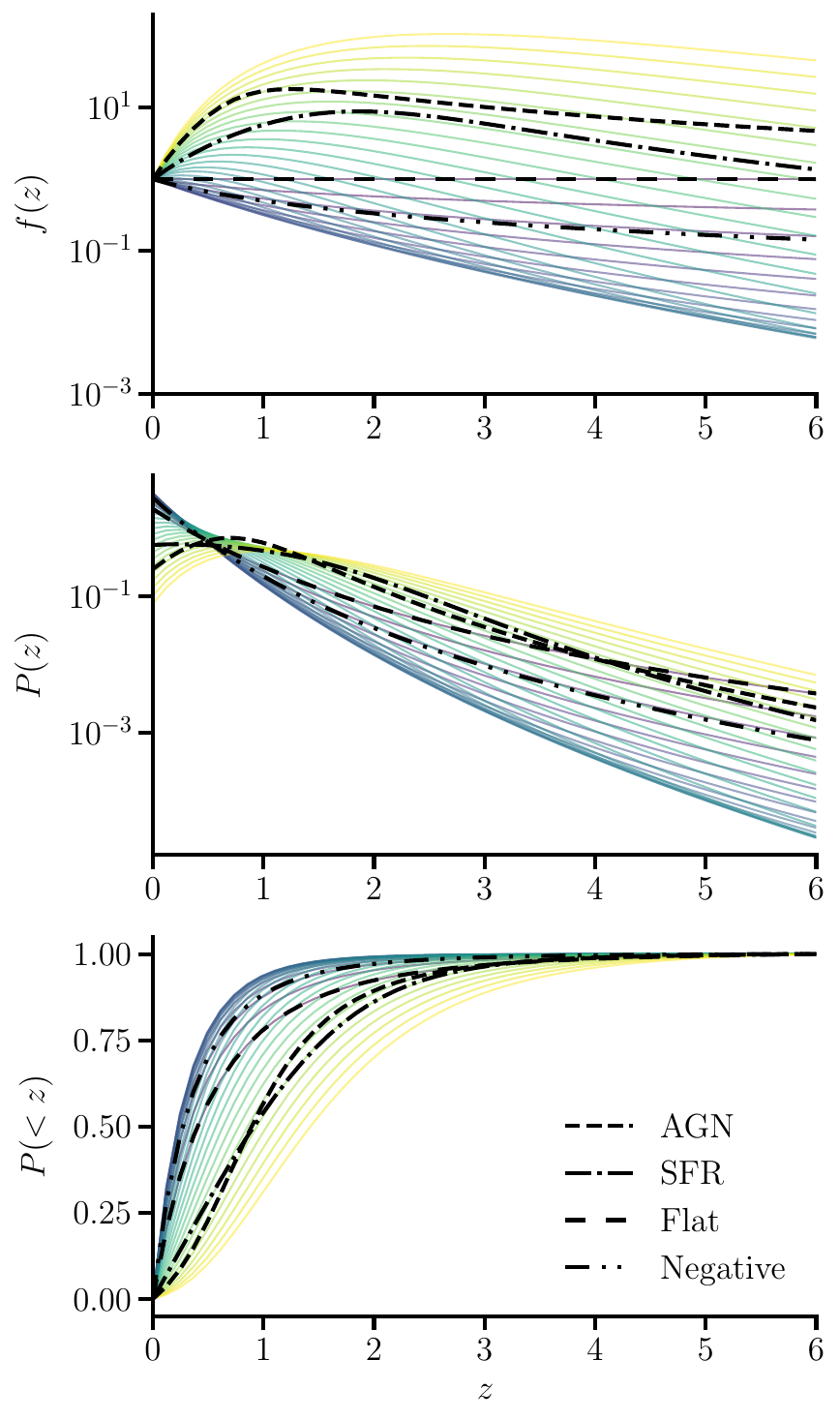}
  \caption{The parameterization for the source density evolution given
    in Equation~\eqref{eqn:source_evolution} is shown as thin colored
    lines for a range of different $p_1$, $p_2$, and $z_c$ values in
    the top panel. Models typically used to represent the evolution
    are also plotted as dashed and dotted lines, scaled to a local
    density of 1 for comparison. The central panel shows the
    differential probability that a neutrino originates at a redshift
    $z$, given that it contributes to the flux at Earth. This
    probability is calculated from the total differential flux above
    $E_\mathrm{min}$, found by evaluating the integrand of
    Equation~\eqref{eqn:total_flux_above_Emin} for the example case of
    $n_0=10^{-6}$~$\mathrm{Mpc}^{-3}$,
    $L=10^{42}$~erg~$\mathrm{s}^{-1}$,
    $E_\mathrm{min} = E_0 = 100$~TeV, $E_\mathrm{max} = \infty$ and
    $\gamma=2$. The bottom panel shows the corresponding cumulative
    probability. The AGN luminosity evolution is taken from the
    best-fit pure luminosity evolution (PLE) model in \citet{Ueda:kv},
    and the SFR evolution is as presented in Equation~(15) of
    \citet{Madau:2014gh}. Flat signifies no evolution and negative is
    $\propto (1+z)^{-1}$.}
  \label{fig:source_evolution}
\end{figure}

As neutrinos propagate from their sources to Earth, they lose energy
due to the adiabatic expansion of the universe such that the arrival
energy is related to the emitted energy at the source by a factor of
$1+z$. In this way, we can write the differential neutrino flux of a
single point source at Earth as
\begin{equation}
  \frac{\mathrm{d}\bar{N}_\nu^\mathrm{src}}{\mathop{\mathrm{d}E} \mathop{\mathrm{d}t} \mathop{\mathrm{d}A}} = L k_\gamma \frac{(1+z)^{2-\gamma}}{4\pi D_L^2(z)} E_0^{-\gamma} \Bigg( \frac{E}{E_0} \Bigg)^{-\gamma}, 
  \label{eqn:point_source_at_Earth}
\end{equation}
where the factor of $(1+z)^{2-\gamma}$ is due to the neutrino energy
losses. This factor arises as we consider $E_\mathrm{min}$ and
$E_\mathrm{max}$, as defined in Equation~\eqref{eqn:luminosity}, to be
the energy range within which the luminosity is defined at Earth, not
the location of the source. We can define the point source flux
normalization, $\phi$, such that
\begin{equation}
  \frac{\mathrm{d}\bar{N}_\nu^\mathrm{src}}{\mathop{\mathrm{d}E} \mathop{\mathrm{d}t} \mathop{\mathrm{d}A}} = \phi \Bigg( \frac{E}{E_0} \Bigg)^{-\gamma},
  \label{eqn:point_source_flux_norm}
\end{equation}
to simplify this expression. The total differential flux can then be
found by integrating over all sources in the population
\begin{equation}
  \frac{\mathrm{d}\bar{N}_\nu^\mathrm{tot}}{\mathop{\mathrm{d}E} \mathop{\mathrm{d}t} \mathop{\mathrm{d}A} \mathop{\mathrm{d}\Omega}}
  = \frac{1}{4 \pi} \int_0^\infty \mathop{\mathrm{d}z}  \frac{\mathrm{d}\bar{N}_\mathrm{s}^\mathrm{tot}}{\mathrm{d}V} \frac{\mathrm{d}V}{\mathrm{d}z}
  \frac{\mathrm{d}\bar{N}_\nu^\mathrm{src}}{\mathop{\mathrm{d}E} \mathop{\mathrm{d}t} \mathop{\mathrm{d}A}}.
  \label{eqn:phi_n}
\end{equation}
Similarly, the total flux of neutrinos between $E_\mathrm{min}$ and
$E_\mathrm{max}$ is given by integrating
Equation~\eqref{eqn:point_source_at_Earth} over energy and then
integrating over the source
population\footnote{Equation~\eqref{eqn:total_flux_above_Emin} has
  been corrected following the erratum published at DOI: \href{https://journals.aps.org/prd/abstract/10.1103/PhysRevD.105.129904}{10.1103/PhysRevD.105.129904}. Fig.~\ref{fig:source_evolution} and the
  rest of the paper are unaffected by this correction and remain as in
  the original published version.}
\begin{equation}
  \frac{\mathrm{d}\bar{N}_\nu^\mathrm{tot}}{\mathop{\mathrm{d}t} \mathop{\mathrm{d}A}} = \int_{0}^{\infty} \mathop{\mathrm{d}z} \int_{E_\mathrm{min}}^{E_\mathrm{max}} \mathop{\mathrm{d}E} \frac{\mathrm{d}\bar{N}_\mathrm{s}^\mathrm{tot}}{\mathrm{d}V} \frac{\mathrm{d}V}{\mathrm{d}z}  \frac{\mathrm{d}\bar{N}_\nu^\mathrm{src}}{\mathop{\mathrm{d}E} \mathop{\mathrm{d}t} \mathop{\mathrm{d}A}}.
  \label{eqn:total_flux_above_Emin}
\end{equation}
For typical source evolutions, as discussed above, we expect the
majority of the neutrinos at Earth to come from considerable distances
($z\sim1$--2), as illustrated in Fig.~\ref{fig:source_evolution}.

\section{\label{sec:nu_detection_icecube}Neutrino detection with
  IceCube}

Neutrinos are detected through their interactions in dense,
transparent media, typically ice or water. In principle, the approach
described in this work can be applied to any neutrino detector, but
here we will focus on the case of the IceCube neutrino
observatory\footnote{\url{https://icecube.wisc.edu}}, given that it is
currently the only instrument to have detected an astrophysical flux
of neutrinos.

The IceCube detector is essentially a cubic-kilometer of ice at the
geographic South Pole that has been instrumented with photomultiplier
detector modules. These modules are installed through a grid of
strings lowered into the ice at regular intervals between depths of
1450~m and 2450~m. Neutrino interactions produce relativistic
secondary particles, which in turn lead to Cherenkov radiation in the
ice. These signals are detected in optical wavelengths by the
photomultiplier modules and are used to identify neutrino-induced
events and reconstruct the direction and energy of these incoming
neutrinos. A detailed description of the IceCube instrumentation can
be found in~\citet{Achterberg:2006kh, Aartsen:2017tl}.

Due to the geometry of the Earth, the energy-dependence of the
neutrino interaction cross-section, and the details of the detection
process, any neutrino detector will have an effective area that is
both a function of the neutrino arrival energy and direction,
$A_\mathrm{eff}(E, \omega)$. As IceCube is located at the South Pole,
its effective area is independent of right ascension and purely a
function of declination, $\delta$. The expected number of detected
neutrinos from a single source can be found by convolving its flux
with the effective area of the detector and integrating over the
exposure time
\begin{equation}
  \bar{N}_\nu^\mathrm{src} = T \int_{E_\mathrm{min}}^{E_\mathrm{max}} \mathop{\mathrm{d}E} A_\mathrm{eff}(E, \delta) \frac{\mathrm{d}\bar{N}_\nu^\mathrm{src}}{\mathop{\mathrm{d}E} \mathop{\mathrm{d}t} \mathop{\mathrm{d}A}},
  \label{eqn:aeff_folding}
\end{equation}
where $T$ is the total observation time. For each event, the
photomultiplier signals are reconstructed to estimate the neutrino
energy and direction. In this way, a resulting dataset is a number of
neutrino events, their reconstructed energies, $\hat{E}$, arrival
directions, $\hat{\omega}$, and the corresponding uncertainties on
these measurements. This information can be used to separate
neutrino-induced events from other background signals. However, it is
more difficult to differentiate between astrophysical neutrinos and
atmospheric neutrinos due to cosmic ray interactions in the Earth's
atmosphere. The origin of a neutrino event can only be determined in a
probabilistic way, with the measured neutrino energy as the driving
factor.

There are two main event types in a neutrino detector like IceCube:
``tracks'' from muons produced in charged-current interactions of muon
neutrinos, and ``cascades'' from charged-current interactions of
electron and tau neutrinos, as well as neutral-current interactions of
all flavors \cite{Gaisser:2016aa}. In this work, we rely on results
from the analysis of up-going muon tracks from the Northern sky,
produced by muon neutrinos interacting both within and outside of the
instrumented volume. In this case, the effective area of the detector
is much larger in comparison with ``cascade''-like neutrino
interactions that must occur within the sensitive volume in order to
be identified. Due to their topology, track events have a better
angular resolution of $\sim0.5^\circ$ at the highest energies, making
them more sensitive to the detection of point sources via the method
described in Section~\ref{sec:detection_probability}. However, only
the energy of the resulting muon is sampled in the detector, so
$\hat{E}$ is a lower limit on the reconstructed muon energy and the
energy resolution is much worse than for cascade events. The geometry
of the Earth and the location of IceCube at the South Pole mean that
events coming from the Northern hemisphere have a lower background,
due to the absorption of atmospheric muons produced in cosmic
ray-induced air showers. The combination of these factors makes the
Northern sky muon track sample the most sensitive to point-like
neutrino sources, and therefore it provides the most substantial
constraints on the source population in our model.

Due to neutrino oscillations, we expect the flavor composition of the
neutrino signal to be equally partitioned at Earth for standard
extra-Galactic source scenarios \cite{Learned:1995ib,
  Athar:2011dm}. As our analysis is based on a muon track sample, we
consider the per-flavor flux summed over muon neutrinos and muon
anti-neutrinos throughout this work.

\section{\label{sec:detection_probability}Individual source detection
  probability}

To understand the implications of a non-detection of sources for our
population model, it is essential to define the criteria required for
detection. Previous work in this area has considered the observation
of neutrino multiplets as a sufficient requirement, where a multiplet
is defined as two or more high energy ($\hat{E} > 100$--200~TeV)
neutrino events that come from similar directions \cite{Lipari:2008ew,
  Ahlers:2014em, Kowalski:2015gm, Murase:2016gly,
  Palladino:2019tm}. While the highest energy neutrinos are the most
likely to be astrophysical and have the best angular resolution, the
atmospheric background is not negligible at these energies, with some
events having a probability of astrophysical origin in the range of
0.5 to 0.7 \cite{Aartsen:2016hp}. Additionally, we do not know a
priori whether two neutrinos originate from the same source. The
angular resolution at these energies is around $0.5^{\circ}$, but the
overlap of two error regions would not be sufficient to report a
detection due to the probability of this occurring by chance, which
becomes increasingly non-negligible as IceCube continues to gather
data.

It is possible to introduce a correction factor for the number of
false multiplets in this approach, as shown in~\cite{Murase:2016gly},
but this implies reducing the constraining power of the analysis. We
also note that in the existing 3-year public dataset
\citep{PSdataset:2018ab}, there are now several
multiplets\footnote{The exact number depends on the declination and
  angular resolution cuts considered.} above 50~TeV, so this criterion
has been surpassed with no reported source detections. In general, the
number of multiplets is expected to increase as more data is
acquired. To continue to use the multiplet approach to define the
non-detection of point sources by IceCube, it is necessary to either
increase the energy threshold or to change the definition to require
three or more events to have consistent directions, again weakening
the subsequent constraints on the properties of the source population.

In this work, we go beyond the assumption of neutrino multiplets
corresponding to a source detection. The IceCube Collaboration has
developed methods for the detection of neutrino sources that are
optimized in terms of their sensitivity to background-dominated
signals. The standard point source search method makes use of the
reconstructed energies and arrival directions of neutrinos with a
likelihood ratio technique for the comparison of source and background
hypotheses, as described in \citet{Braun:2008kr}. In order to include
all relevant information into the likelihood and maximize the
discovery potential, these searches also make use of lower energy
events in the sample down to $\sim100$~GeV. As a consequence, the
IceCube point source search is more sensitive to source detection than
solely considering high-energy events. A more sensitive detection
criterion implies more stringent constraints on the source population
for a non-detection, and is therefore a favorable starting point for
this analysis. The point source search also has the benefit of
consistently accounting for statistical fluctuations and the presence
of the atmospheric neutrino background. We briefly describe the
IceCube point source search methodology in
Section~\ref{sec:IceCube_PS}, then explain how we can use this
analysis to define the detection probability in
Section~\ref{sec:pdet_def}.

\subsection{\label{sec:IceCube_PS}The IceCube point source search}

The unbinned likelihood ratio method used by the IceCube Collaboration
is based on the likelihood function for the reconstructed neutrino
energies and arrival directions at a general, test source point on the
sky, $\omega_s = (\alpha_s, \delta_s)$. The likelihood is essentially
a mixture model over the possible source and background contributions
with the form
\begin{equation}
  \mathcal{L}(\omega_s, \bar{N}_\nu^\mathrm{src}, \gamma) = \prod_{i = 1}^{N_\nu}  \Bigg[ \frac{\bar{N}_\nu^\mathrm{src}}{N_\nu} S_i (\omega_s, \gamma)+ \bigg(1 - \frac{\bar{N}_\nu^\mathrm{src}}{N_\nu}\bigg)B_i \Bigg] P(\gamma),
  \label{eqn:point_source_likelihood}
\end{equation}
where $N_\nu$ is the total number of neutrinos in the sample,
$\bar{N}_\nu^\mathrm{src}$ is the expected number of signal events,
and $S_i$ and $B_i$ are the signal and background likelihoods
respectively for the $i^\mathrm{th}$ event. There is also an optional
prior term, $P(\gamma)$, to include information on the spectral index
from analyses of the diffuse flux. The source and background
likelihoods factor into independent spatial and energy likelihoods for
the event properties
\begin{equation}
  S_i = P(\hat{\omega} | \omega_s, \sigma_i)P(\hat{E} | \gamma),
\end{equation}
\begin{equation}
  B_i = P(\hat{\omega}| \mathrm{BG})P(\hat{E} | \mathrm{BG}),
\end{equation}
where $\sigma_i$ is the event-by-event uncertainty on the angular
reconstruction. In practice, events far from $\omega_s$ have a
negligible probability of contributing to the likelihood, so a band in
declination is selected for a more efficient analysis. In this case,
$N_\nu$ is the total number of events in the band, and the background
spatial likelihood for isotropic emission is a function of the
declination-dependence of the atmospheric neutrino background in the
band~\citep{Aartsen:2015dc}.

A set of $\omega_s$ is then defined to scan over the sky. At each
source position, the likelihood is maximized with respect to
$\bar{N}_\nu^\mathrm{src} \geq 0$ and $1 \leq \gamma \leq 4$. The
best-fit $\hat{N}_\nu^\mathrm{src}$ and $\hat{\gamma}$ are used in a
likelihood ratio test to compare the hypotheses of clustered point
source emission and isotropic background. The likelihood ratio test
statistic is
\begin{equation}
  \lambda = -2 \log \Bigg[ \frac{P(\omega_s, N_\nu^\mathrm{src}=0 | \hat{\omega}, \hat{E})}{P(\omega_s, \hat{N}_\nu^\mathrm{src}, \hat{\gamma} | \hat{\omega}, \hat{E})} \Bigg].
\end{equation}

To understand the performance of the analysis and the significance of
the results, the distribution of $\lambda$ is profiled through Monte
Carlo simulations for the case of purely background events. The
$\lambda$ corresponding to the $5\sigma$ tail probability,
$\lambda_{5\sigma}$, can then be defined, which is typically between
20 and 30. By simulating the $\lambda$ distribution in the case of
injected source events, it is also possible to define the ``discovery
potential,'' $\phi_\mathrm{dp}$, as the point source flux which leads
to $>5\sigma$ deviation from background for 50\% of repeated simulated
trials, given $\omega_s$ and $\gamma$. As a scan of point source
searches is carried out at different locations over the sky, a source
found to have a local significance of $5\sigma$ would then have a
trial factor correction applied to calculate the final p-value
relevant for a discovery announcement. This is estimated in various
ways from simulated experiments, and typically has a value of
$\sim 10^5$, such that a source with a pre-trial p-value of $10^{-7}$
would have a post-trial p-value of only $10^{-2}$.

\subsection{\label{sec:pdet_def}Defining the detection probability}

We can make use of the IceCube point source search framework in order
to quantify source detection. We define the requirement for the
detection of a neutrino point source as $\lambda >
\lambda_{5\sigma}$. While this is not the criterion that would be used
to announce a discovery due to the lack of a trial factor correction,
it is the most constraining criterion that can be used to connect with
the current results of the IceCube point source searches. In this way,
the detection probability of a source will depend on its flux, its
spectral index, and its location on the sky in relation to IceCube's
effective area and so have the form
\begin{equation}
  P_\mathrm{det} = P(\mathrm{det} | \phi, \gamma, \delta),
  \label{eqn:detection_probability}
\end{equation} 
where we have used the fact that IceCube's effective area is
independent of right ascension and used the notation for the point
source flux introduced in Equation~\eqref{eqn:point_source_flux_norm}.

\begin{figure}[t]
  \centering
  \includegraphics[width=\columnwidth]{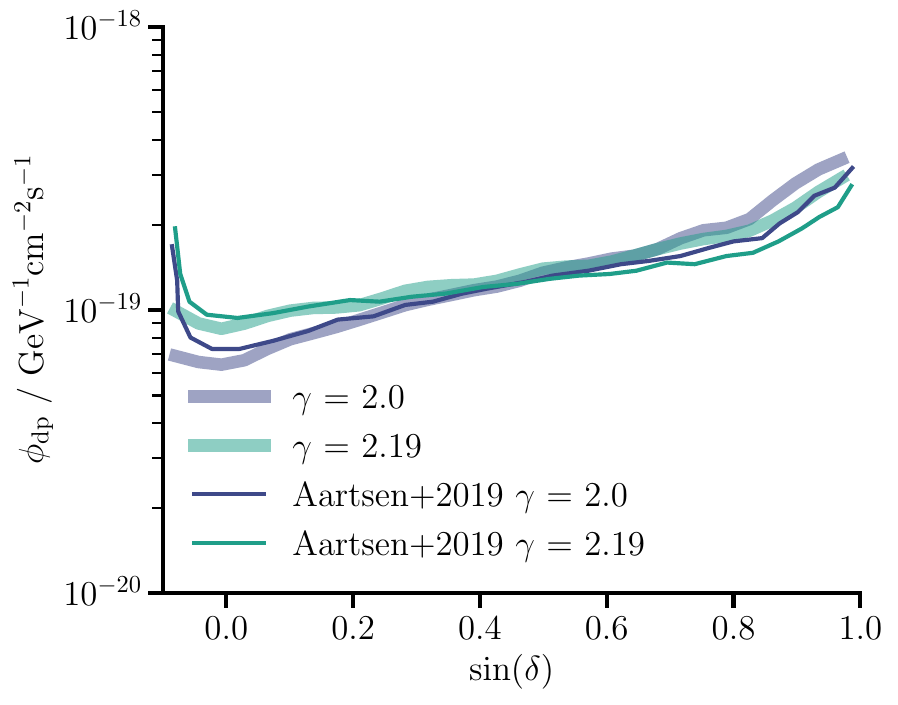}
  \caption{The discovery potential calculated using the method
    described in Section~\ref{sec:IceCube_PS} applied to simulated
    data in order to reproduce the discovery potential presented in
    Fig. 18 of \citet{Collaboration:2019db}, which are shown for
    comparison. The flux is shown normalized to $E_0 = 100$~TeV. }
  \label{fig:recreate_Aartsen+2019}
\end{figure}

The IceCube point source analyses usually report $\phi_\mathrm{dp}$ as
a function of declination for specific spectral indices, e.g., Fig.~3
and Fig.~18 in \citet{Collaboration:2019db}, as well as
Fig.~\ref{fig:recreate_Aartsen+2019} here. These plots directly give
the source parameters corresponding to $P_\mathrm{det} =
0.5$. However, to fully characterize how the detection probability
depends on the source parameters, we use simulations to profile the
$\lambda$ distribution for a range of relevant source positions and
properties. In our simulations, we make use of the publicly available
information regarding the IceCube effective area to muon track events
from~\citet{PSdataset:2018ab} to calculate the total number of
neutrinos for a given flux. We also use the angular resolution
information reported in~\citet{PSdataset:2018ab} to model the
distribution of the reconstructed arrival direction $\hat{\omega}$,
and the correspondence between reconstructed muon energy and true
neutrino energy reported in~\citet{Aartsen:2015de} to estimate the
energy resolution and model the distribution of the reconstructed
energy~$\hat{E}$. We approximate the atmospheric neutrino spectrum as
a broken power law, and the astrophysical spectrum is modeled with a
single power law. To evaluate the discovery potential, we combine our
simulated datasets with the procedure outlined in
Section~\ref{sec:IceCube_PS}. We verify the ability of this approach
to recreate the reported discovery potential plots from existing
IceCube publications, showing our results in
Appendix~\ref{sec:Pdet_Braun}.

We then implement our approach to reproduce the sensitivity of the
analysis described in \citet{Collaboration:2019db}. To do this, we
follow their analysis by simulating 497,000 events to reflect eight
years of data, and incorporating the results of \citet{Aartsen:2016hp}
and \citet{Haack:2017ab} by modeling a diffuse astrophysical
background as well as the atmospheric component. We also include a
prior on the spectral index. Due to the improved discovery potential
of including lower energy events into the analysis, we follow the
approach of IceCube and assume that sources have a power-law neutrino
spectrum that extends to energies as low as
$E_\mathrm{min}^\mathrm{PS} = 1$~TeV and up to
$E_\mathrm{max}^\mathrm{PS} = 100$~PeV when calculating the detection
probability used in this work. The minimum energy is slightly higher
than the lowest energies considered for the sample used in
\citet{Collaboration:2019db} (see Table~1 therein) as it was not
possible to estimate the energy resolution of the detector well below
1~TeV using the publicly available information reported
in~\citet{Aartsen:2015de}. However, the impact of this on the
presented detection probability is negligible given the small
effective area of IceCube below 1~TeV and the relative contribution of
the lowest energy events in the sample to the point source likelihood
function described in
Equation~\eqref{eqn:point_source_likelihood}. Taking these factors
into account, we can reasonably reproduce the discovery potential as a
function of declination, as shown in
Fig.~\ref{fig:recreate_Aartsen+2019}. The overall shape is well
reproduced for the different spectral indices, but we see that at
higher $\sin(\delta)$, the discovery potential is slightly
overestimated. This is likely due to the approximate treatment of the
effective area and instrument response that we have used in our
simulations, based on publicly available information. In contrast, the
IceCube Collaboration analyses make use of much more lower-level
information and detailed simulations. In any case, the impact of
overestimating the discovery potential at this level is negligible on
the final results.

\begin{figure}[h!]
  \centering
  \includegraphics[width=0.9\columnwidth]{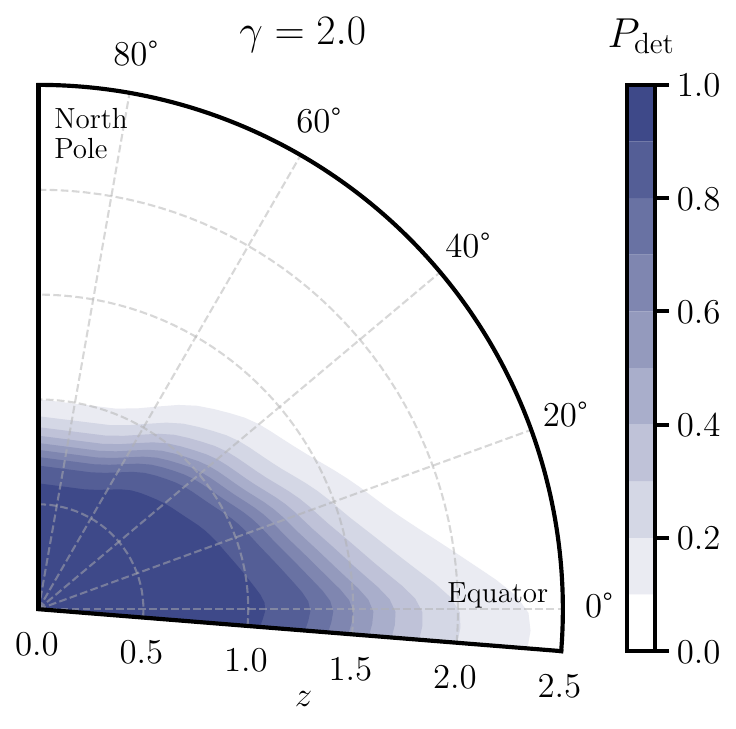}
  \includegraphics[width=0.9\columnwidth]{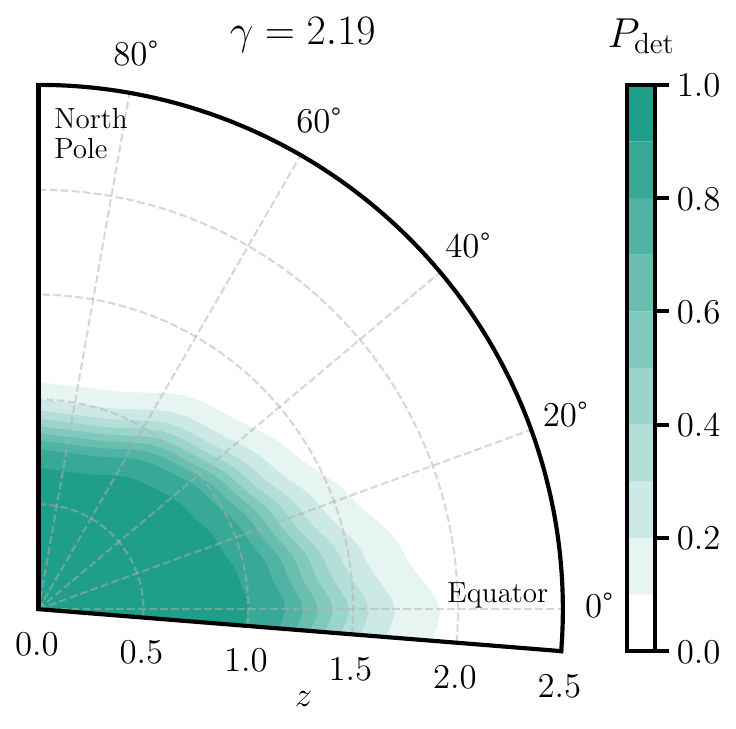}
  \caption{The detection probability is shown as a function of
    redshift (radial axis), declination (angular axis), and spectral
    index (upper and lower panels). The weight of the colored segments
    shows the detection probability from 0 to 1, as indicated by the
    color bars. For simplicity, we show ten detection probability bins
    and two spectral indices. In practice, we can compute these values
    with finer resolution, as required. In both panels, the detection
    probability is calculated for a source with a constant emission
    rate above 100~TeV of
    $\mathrm{d}N_\nu/\mathrm{d}t = 10^{44}$~$\mathrm{s}^{-1}$ to
    facilitate comparison. Here, 100~TeV is chosen as an example and
    the actual calculation of the detection probability includes
    neutrinos with energies down to 1~TeV, as detailed in the text. At
    high declinations, the results are extrapolated from
    $\sin(\delta) = 0.97$ to avoid boundary effects.}
  \label{fig:Pdet_multi}
\end{figure}

Having verified this method, we can then use it to calculate
$P(\mathrm{det | \phi, \gamma, \delta})$ for a range of source
parameters, as shown in Fig.~\ref{fig:Pdet_multi}. We see that for
harder spectral indices, the detection probability is highly
declination dependent. This variation is because of the energy
dependence of the IceCube effective area, and the importance of high
energy neutrinos in discriminating potential sources from the
background. At high declinations, the highest energy neutrinos suffer
Earth-absorption and are not detected. Conversely, at low
declinations, high energy source neutrinos can make a substantial
contribution to the signal. For this reason, a source producing a
constant number of neutrinos above a certain energy can be detected
from higher redshift if it has a harder spectral index and is located
at lower declinations.

The detection probability can be included in the physical picture
described in Section~\ref{sec:physical_model} by defining the expected
number of detected sources in a given population as
\begin{equation}
  \bar{N}_\mathrm{s} = \frac{1}{2} \int_{\delta_\mathrm{min}}^{\pi/2} \mathop{\mathrm{d}\delta} \cos \delta \int_0^{\infty} \mathop{\mathrm{d}z} P(\mathrm{det} | \phi, \gamma, \delta) \frac{\mathrm{d}\bar{N}_\mathrm{s}^\mathrm{tot}}{\mathrm{d}V}\frac{\mathrm{d}V}{\mathrm{d}z},
  \label{eqn:expected_sources}
\end{equation}
where $\delta_\mathrm{min} \simeq -5^\circ$ is the minimum declination
used in the search for sources, given that we focus on Northern sky
searches, and we note that $\phi$ is a function of $z$, as shown in
Equations~\eqref{eqn:point_source_at_Earth}~and~\eqref{eqn:point_source_flux_norm}.

\subsection{\label{subsec:Potential issues}Limitations of this
  approach}

The IceCube approach to defining the discovery potential implicitly
relies on inevitable stopping and testing intentions that must be
defined when calculating p-values \cite{Kruschke:2017kw}. This
strategy means that a set number of neutrinos are used in the
background and injected source simulations, reflecting the observed
number of neutrinos in the point source analysis dataset. This
procedure connects the individual source detection probability to the
higher-level population parameters through the source flux,
$\phi(L, z)$, and spectral index, $\gamma$. However, we also need to
account for the effect of varying $n_0$ and $\theta$. Changing the
source density and evolution will lead to different diffuse
astrophysical neutrino backgrounds, and therefore different background
$\lambda$ profiles and associated $\lambda_{5\sigma}$ levels. In this
way, the full detection probability has the form
\begin{equation}
  P_\mathrm{det} = P(\mathrm{det} | \phi, \gamma, \delta, n_0, \theta).  
\end{equation}
Calculating the detection probability as a function of these
parameters is very computationally expensive, as noted in Section~5 of
\citet{Collaboration:2019db}, in which this has been done for fixed
values of $\gamma$ and $\theta$. Additionally, as the diffuse
background to point source detection is dominated by the atmospheric
component, we do not expect a significant impact of this effect on the
results, as quantified in Appendix~\ref{sec:atmosonly_Pdet}. For this
reason, we use the implementation of the detection probability
described in Equation~\eqref{eqn:detection_probability} in this work.

\section{\label{sec:statistical_model}Statistical formalism}

The physical model presented in Section~\ref{sec:physical_model} and
the detector description given in
Section~\ref{sec:nu_detection_icecube} connect the high-level
population parameters to the detected neutrinos and their
properties. This dataset can then be used to reconstruct the observed
astrophysical flux and to search for individual sources, as detailed
in Section~\ref{sec:detection_probability}. In order to use this
connection to place constraints on the nature of the sources, we need
to derive a likelihood function for these observations, given the
model assumptions and parameters. The case of observations of a
population of similar sources is particularly well-suited to a
statistical framework known as hierarchical or multi-level modeling
\cite{Gelman:2014aa}, where the individual sources share certain
properties. We use a Bayesian approach to infer the allowed regions of
parameters through the computation of the posterior distribution,
under the assumptions of our model and the observations.

Ideally, the likelihood function would be derived for the lowest level
available data, the individual neutrino energies, and arrival
directions. However, given the computational complexity of this
implementation and the limited publicly available information, we
choose to take a more practical approach that makes use of existing
IceCube results. We first introduce the observable quantities used in
Section~\ref{subsec:observable_quantities} and then make use of the
hierarchical modeling approach to construct the likelihood function in
Section~\ref{sec:hierarchical_model}, along with a discussion of the
parameter priors chosen. The computational implementation of this
likelihood in order to perform inference is described in
Section~\ref{sec:implementation}.

\subsection{\label{subsec:observable_quantities}Reconstructed
  quantities}

The IceCube collaboration measures the flux of astrophysical neutrinos
using a maximum likelihood analysis based on the reconstructed event
energies and arrival directions. The astrophysical neutrino spectrum
is typically reported in the form
\begin{equation}
  \frac{\mathrm{d}N_\nu^\mathrm{tot}}{\mathop{\mathrm{d}E} \mathop{\mathrm{d}t} \mathop{\mathrm{d}A} \mathop{\mathrm{d}\Omega}} = (\hat{\Phi}\pm \sigma_\Phi) \Bigg(\frac{E}{E_0}\Bigg)^{-(\hat{\gamma}\pm\sigma_\gamma)},
  \label{eqn:obs_spec}
\end{equation}
where $E_0$ is 100~TeV, and $\hat{\Phi}$ is the estimated differential
flux normalization at $E_0$. The uncertainties on the reconstructed
flux and spectral index are given by $\sigma_\Phi$ and
$\sigma_\gamma$, respectively. These represent the 68\% confidence
intervals based on the profile likelihood using Wilk's theorem
\cite{Wilks:1938vs} and include both statistical and systematic
uncertainties. Using the physical model described in
Section~\ref{sec:physical_model}, we can express the total
differential flux at Earth by integrating over the contribution due to
all sources in the population, as shown in Equation~\eqref{eqn:phi_n}.

The detection probability of a single source is defined in
Section~\ref{sec:detection_probability}. Equation~\eqref{eqn:expected_sources}
represents the number of sources that we would expect to detect from a
population, given the population parameters and the IceCube
sensitivity of the point source analysis. The non-observation of point
sources in this framework can be summarized by $N_\mathrm{s} = 0$
given $P(\mathrm{det} | \phi, \gamma, \delta)$, where $N_\mathrm{s}$
is the detected number of sources. This result can then be used to
quantify the implications for the population model.

\subsection{\label{sec:hierarchical_model}Hierarchical model}

\begin{figure}[t]
  \centering \includegraphics[width=0.9\columnwidth]{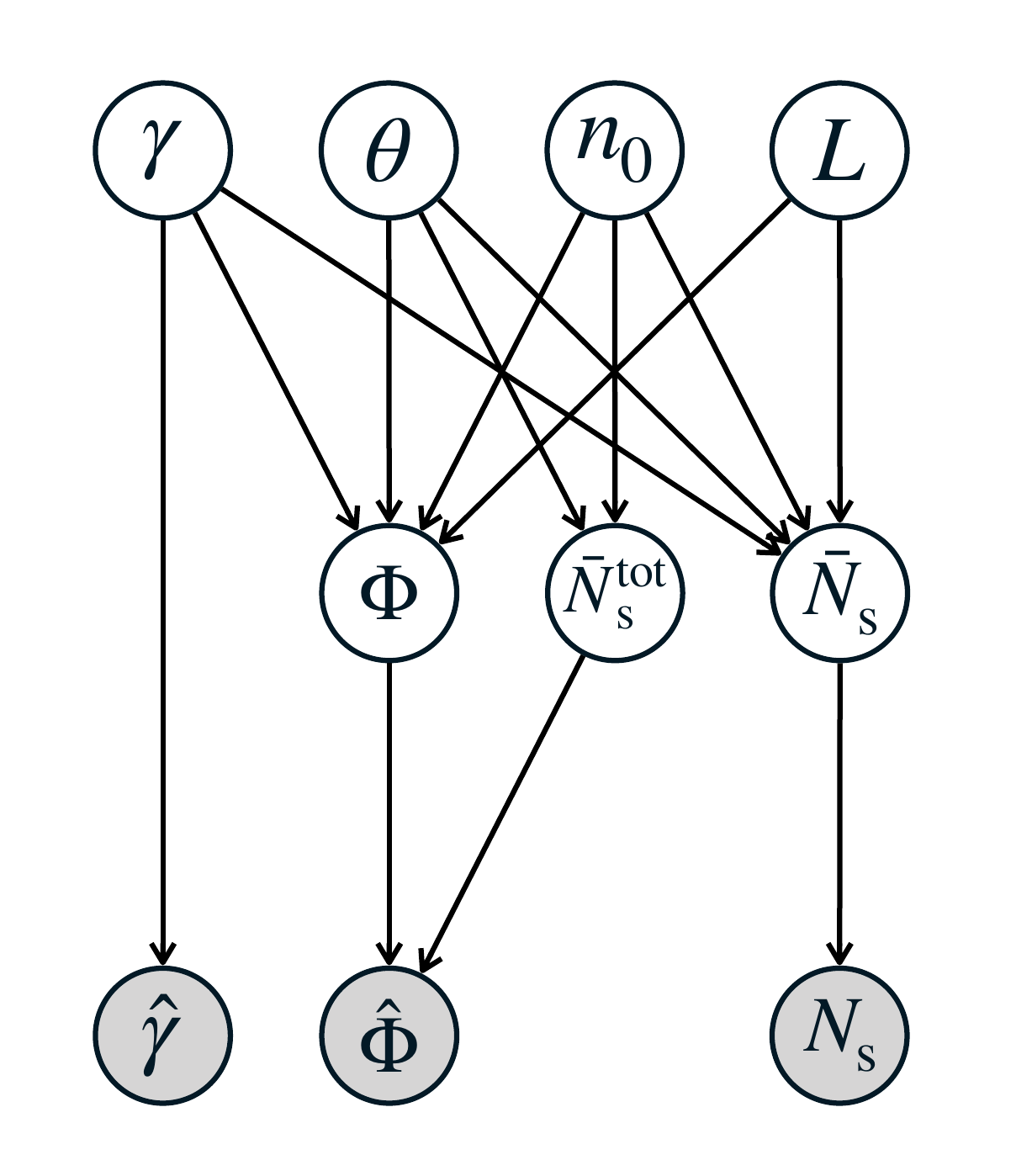}
  \caption{Directed acyclic graph showing the hierarchical likelihood
    function given in Equation~\eqref{eqn:likelihood}. The unshaded
    circles show the free parameters, and the shaded circles represent
    observed quantities, with the arrows showing the connections
    between them. The parameters are described in the text.}
  \label{fig:dag}
\end{figure}

We derive an expression for the unnormalized posterior distribution
through a hierarchical modeling approach. The idea is to connect the
constraints on the model parameters by requiring our population model
to explain both the diffuse flux observations and the non-detection of
point sources. The structure of the likelihood is shown
diagrammatically in Fig.~\ref{fig:dag}. This figure can be understood
as a representation of the generative model needed to produce the
IceCube observations. If we were to simulate these observations, we
would start by defining the top-level parameters, use these to derive
the lower level or ``latent'' parameters, and finally sample the
observations based on the values of these latent parameters. Writing
down the generative model for the observations gives us all the
ingredients needed to derive the likelihood function. The full
derivation for the posterior distribution is given in
Appendix~\ref{sec:full_posterior_derivation}. Here, we use
Fig.~\ref{fig:dag} to intuitively state the connections between
observations and parameters and then discuss the form of the resulting
expressions. The model parameters are summarized in
Table~\ref{tab:parameters} for reference.

\begin{table}[ht]
  \centering
  \begin{tabular}{ll}
    \toprule
    \textbf{Parameters} & \\ \midrule
    $\gamma$ & Spectral index \\
    $\theta$ & Population evolution shape $\{p_1, p_2, z_c\}$ \\
    $n_0$ & Local source density \\
    $L$ & Observed muon neutrino luminosity \\
    $\Phi$ & Diffuse flux normalization \\
    $\bar{N}_\mathrm{s}^\mathrm{tot}$ & Expected number of sources in the universe \\
    $\bar{N}_\mathrm{s}$ & Expected number of detected sources \\ \midrule
    \multicolumn{2}{l}{\textbf{Observables}} \\ \midrule
    $\hat{\gamma}$ & Reconstructed $\gamma$ \\
    $\hat{\Phi}$ & Reconstructed $\Phi$ \\
    $N_\mathrm{s}$ & Number of detected sources \\
    \bottomrule
  \end{tabular}
  \caption{Summary of model parameters and observables. The luminosity
    is defined between $E_\mathrm{min}^L = 10$~TeV and
    $E_\mathrm{max}^L = 10$~PeV and the diffuse flux normalisation is
    also at $E_0 = 100$~TeV.}
  \label{tab:parameters}
\end{table}

\begin{figure}[h!]
  \centering
  \includegraphics[width=\columnwidth]{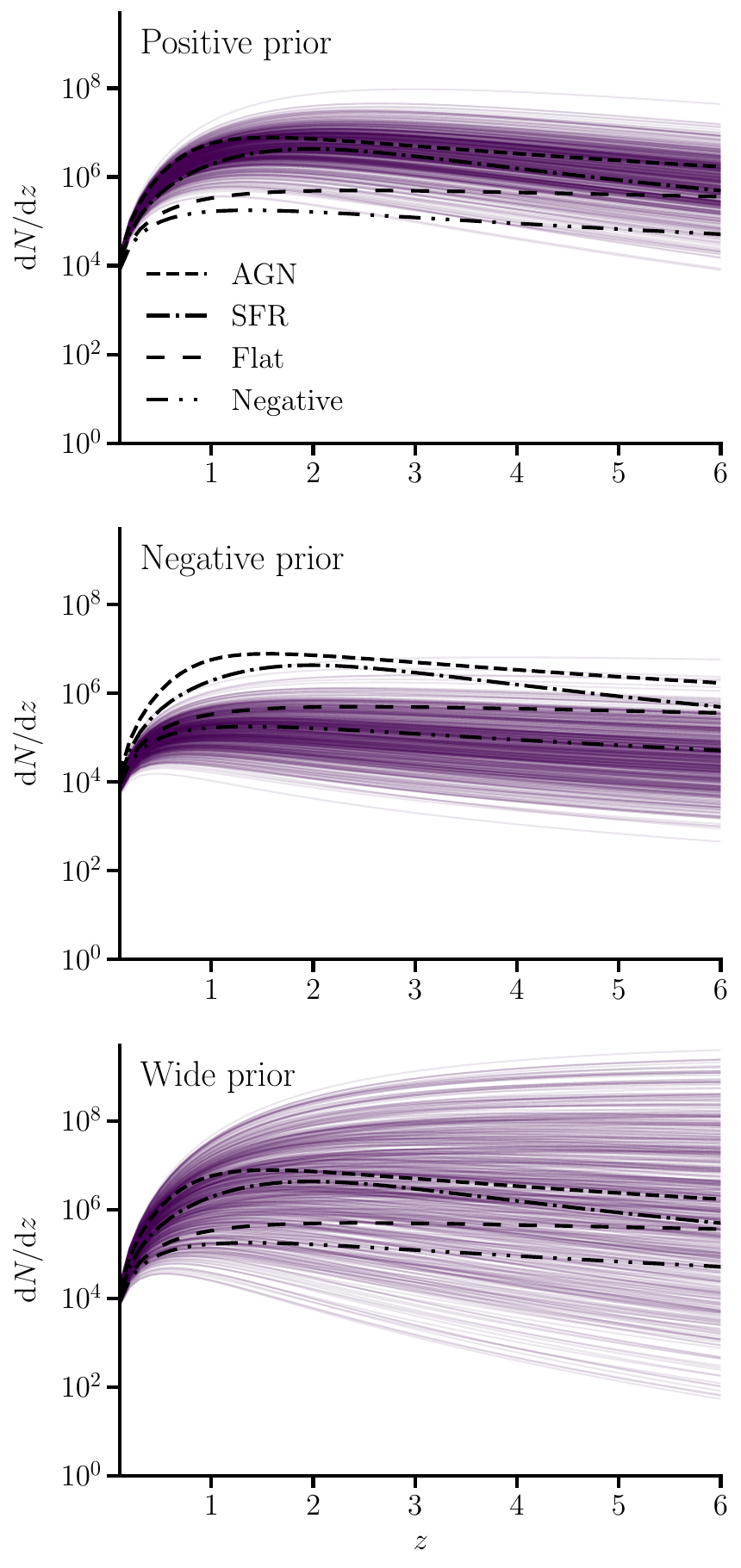}
  \caption{The source redshift distribution is shown for the three
    different priors on the source evolution parameters,
    $\theta = \{p_1, p_2, z_c \}$, as presented
    in~Table~\ref{tab:priors}. The top panel shows the case where the
    prior on $\theta$ is chosen to reflect a positively evolving
    source distribution, similar to that of the SFR or AGN luminosity
    function, the middle panel shows a flat or negative evolution, and
    the bottom panel shows a relatively unconstrained evolution
    model. In each case, 500 random draws from the prior are shown, so
    the density of curves reflects the weight of the prior
    probability. Typical source evolution choices are shown for
    comparison, as in Fig.~\ref{fig:source_evolution}.}
  \label{fig:theta_prior}
\end{figure}

Starting at the top of Fig.~\ref{fig:dag}, the high-level parameters
or hyperparameters are assumed to be independent. To include knowledge
on our modeling assumptions and avoid giving weight to unphysical
regions of the parameter space, weakly informative prior
distributions~\cite{Gelman:2014aa, Simpson:2017kw} are placed on the
high-level parameters, with $n_0$ and $L$ described by broad lognormal
distributions to reflect our uncertainty, and bounded to sensible
values. The evolution parameters, $\theta$, can either be left free to
cover a range of plausible source evolutions or more constrained to
follow positive or negative evolution models, as shown in
Fig.~\ref{fig:theta_prior}. We summarize the high-level parameters and
their priors in Table~\ref{tab:priors}. In both
Fig.~\ref{fig:theta_prior} and Table~\ref{tab:priors}, we refer to the
priors on $\theta$ as ``positive'', ``negative'' and ``wide'', where
the first two terms reflect the source evolution models conisdered and
the latter reflects the choice of a wide range of possible
evolutions. Stepping down one level in Fig.~\ref{fig:dag}, we have the
latent parameters, $\Phi$, $\bar{N}_\mathrm{s}^\mathrm{tot}$, and
$\bar{N}_\mathrm{s}$. These latent quantities can all be directly
calculated from the high-level model parameters, as described by
Equations~\eqref{eqn:phi_n}, \eqref{eqn:total_sources}, and
\eqref{eqn:expected_sources}, respectively.

Finally, the likelihood for the observations is given by
\begin{equation}
  P(\hat{\gamma}, \hat{\Phi}, N_\mathrm{s} | \bar{N}_\mathrm{s}, \bar{N}_\mathrm{s}^\mathrm{tot}, \Phi) = P(\hat{\gamma} | \gamma) P(\hat{\Phi} | \bar{N}_\mathrm{s}^\mathrm{tot}, \Phi) P(N_\mathrm{s} | \bar{N}_\mathrm{s}),  
  \label{eqn:likelihood}
\end{equation}
where the conditional terms connect the high-level parameters to the
observations through the latent parameters. $P(\hat{\gamma} | \gamma)$
is approximated by a normal distribution with mean $\gamma$ and
standard deviation $\sigma_\gamma$. Similarly,
$P(N_\mathrm{s} | \bar{N}_\mathrm{s})$ is given by a Poisson
distribution with mean $\bar{N}_\mathrm{s}$. In the high $n_0$ limit,
the final term $P(\hat{\Phi} | \bar{N}_\mathrm{s}^\mathrm{tot}, \Phi)$
is effectively independent of $\bar{N}_\mathrm{s}^\mathrm{tot}$ and
lognormal. However, in the low $n_0$ limit, the observed astrophysical
flux depends on the presence of sources in the observable universe,
and a universe with no sources is incompatible with the observation of
an astrophysical flux. We can encode this by expanding this term as a
mixture model over the two relevant cases
\begin{equation}
  \begin{split}
    P(\hat{\Phi} | \bar{N}_\mathrm{s}^\mathrm{tot}, \Phi) = & \
    P(N_\mathrm{s}^\mathrm{tot}=0 |
    \bar{N}_\mathrm{s}^\mathrm{tot})P(\hat{\Phi} |
    N_\mathrm{s}^\mathrm{tot} = 0) \\
    & + P(N_\mathrm{s}^\mathrm{tot} \neq 0 |
    \bar{N}_\mathrm{s}^\mathrm{tot})P(\hat{\Phi} | \Phi).
  \end{split}
  \label{eqn:mixture_model}
\end{equation}
where the first term in the mixture is the Poisson probability of
observing $N_\mathrm{s}^\mathrm{tot} = 0$, given
$\bar{N}_\mathrm{s}^\mathrm{tot}$, multiplied by a lognormal
distribution for $\hat{\Phi}$, centered on an effectively zero flux
with standard deviation $\sigma_\Phi$. Similarly, the second term is
weighted by the Poisson probability that
$N_\mathrm{s}^\mathrm{tot} \neq 0$ multiplied by a lognormal
distribution with mean $\Phi$ and standard deviation $\sigma_\Phi$.

\begin{table}[ht]
  \centering
  \begin{tabular}{ccccc}
    \toprule
    & \textbf{Prior} & \textbf{Lower} & \textbf{Upper} & \textbf{Unit} \\ \midrule
    $n_0$ & $\mathrm{Log\mathcal{N}}(\log(10^{-6}), 30)$ & $10^{-12}$ & $10^{2}$ & Mpc$^{-3}$  \\ 
    $L$ & $\mathrm{Log\mathcal{N}}(\log(10^{42}), 30)$ & $10^{35}$  & $10^{50}$ & TeV s$^{-1}$ \\ 
    $\gamma$ & $\mathcal{N}(2, 3)$ & 1  & 4 & - \\ \midrule
    \multicolumn{5}{c}{$\theta$ -- positive} \\ \midrule
    $p_1$ & $\mathcal{N}(19.3, 0.2)$ & 18.1 & 20.5 & - \\
    $p_2$  & $\mathcal{N}(24.9, 0.2)$ & 23.7 & 26.1 & - \\
    $z_c$ & $\mathcal{N}(1.76, 0.12)$ & 1.04 & 2.48 & - \\ \midrule 
    \multicolumn{5}{c}{$\theta$ -- negative} \\ \midrule
    $p_1$ & $\mathcal{N}(19.3, 0.2)$ & 18.1 & 20.5 & - \\
    $p_2$  & $\mathcal{N}(20.5, 0.2)$ & 19.3 & 21.7 & - \\
    $z_c$ & $\mathcal{N}(1.0, 0.08)$ & 0.52 & 1.48 & - \\ \midrule
    \multicolumn{5}{c}{$\theta$ -- wide} \\ \midrule
    $p_1$ & $\mathcal{U}(19, 21)$ & 19 & 21 & - \\
    $p_2 - p_1$  & $\mathcal{U}(0, 6)$ & 0 & 6 & - \\
    $z_c$ & $\mathcal{U}(1.0, 1.8)$ & 1.0 & 1.8 & - \\ \midrule
    $M$ & \multicolumn{4}{p{6.5cm}}{Represents other prior assumptions related to the choice of the physical model.} \\ 
    \bottomrule
  \end{tabular}
  \caption{Prior choices for the high-level model parameters. The
    columns labeled ``Lower'' and ``Upper'' given the lower and upper
    bounds placed on parameters. The three different cases for the
    source evolution parameters, $\theta$, correspond to the three
    different panels in Fig.~\ref{fig:theta_prior}. We also introduce
    $M$ to encode the implicit conditioning of our results on the
    choice of an astrophysical source population made up of discrete
    sources with power-law spectra and shared luminosities. The
    notation $\mathrm{Log}\mathcal{N}$(mean, standard deviation),
    $\mathcal{N}$(mean, standard deviation) and $\mathcal{U}$(lower
    bound, upper bound) is used to summarize the lognormal, normal and
    uniform distributions, repectively. \label{tab:priors}}
\end{table}

\subsection{\label{sec:implementation}Inference}

For a given set of observations, we infer the model parameters using
Bayesian inference. Bringing together the likelihood function from
Equation~\eqref{eqn:likelihood} and the priors described in
Table~\ref{tab:priors}, the joint posterior distribution for the model
parameters, given the observations is
\begin{equation}
  \begin{split}
    P(\theta, n_0, \gamma&, \bar{N}_\mathrm{s}, \bar{N}_\mathrm{s}^\mathrm{tot}, \Phi | N_\mathrm{s}, \hat{\Phi}, \hat{\gamma}, M) \\
    \propto&~P(\theta, n_0, L, \gamma | M) P(\bar{N}_\mathrm{s} | \theta, n_0, L, \gamma) \\
    & \times P(\bar{N}_\mathrm{s}^\mathrm{tot} | n_0, \theta) P(\Phi | \theta, n_0, L, \gamma) \\
    & \times P(\hat{\gamma} | \gamma) P(N_\mathrm{s} |
    \bar{N}_\mathrm{s}) P(\hat{\Phi} |
    \bar{N}_\mathrm{s}^\mathrm{tot}, \Phi).
  \end{split}
  \label{eqn:posterior}
\end{equation}
The full derivation of this expression for the posterior distribution
is given in Appendix~\ref{sec:full_posterior_derivation}. We compute
the posterior distribution numerically by drawing samples from it
using \texttt{Stan}~\cite{Carpenter:2017ke}. This procedure allows us
to easily marginalize over the source evolution parameters, the
spectral index, and the latent parameters of the model to produce the
results shown in Section~\ref{sec:results}. \texttt{Stan} is an
open-source framework for statistical computation with an interface to
several different techniques. In particular, we make use of
\texttt{Stan}'s adaptive Hamiltonian Monte Carlo
\cite{Betancourt:2017kb}. To assess the convergence of the resulting
chains to the target distribution, we employ a set of diagnostics
implemented within the \texttt{Stan} framework. We require that all
chains are sampled with no divergent transitions. Also, for each
parameter, we ensure that we have a large effective sample size,
$n_\mathrm{eff} > 1000$ and that the Gelman-Rubin statistic is
bounded, $\hat{R} < 1.1$, in order to monitor the mixing between
chains.

Equations~\eqref{eqn:phi_n} and \eqref{eqn:expected_sources} contain
integrals over $z$ from 0 to $\infty$. In practice, we integrate up to
some finite $z_\mathrm{max}$ to describe the whole Universe, as
sources beyond a redshift of $\sim6$ are expected to have a negligible
contribution to the flux at Earth for typical source evolutions, as
illustrated in Fig.~\ref{fig:source_evolution}. We verify that the
results are not sensitive to the choice of $z_\mathrm{max}$. We
include the detection probability by fitting a sigmoid over $\phi$ and
then tabulating the parameters of the sigmoid over $\gamma$ and
$\delta$. These values are then interpolated over in the model in the
calculation of $\bar{N}_\mathrm{s}$.

\section{\label{sec:application}Application to IceCube observations}

We apply the statistical formalism developed in
Section~\ref{sec:statistical_model} to infer constraints on the
steady-state source population parameters from the combined
observation of an astrophysical flux but no individual sources. For
the non-observation of point sources, we use the results of the
point-like source search presented in
\citet{Collaboration:2019db}. These results are implemented as
$N_\mathrm{s} = 0$, given that the detection probability is as defined
in Section~\ref{sec:detection_probability}, which is not satisfied by
any direction in the all-sky analysis in \citet{Collaboration:2019db}
(see e.g., Fig. 5 therein). \citet{Collaboration:2019db} make use of
the same muon neutrino sample developed for the diffuse astrophysical
flux analysis of \citet{Aartsen:2016hp} and \citet{Haack:2017ab},
expected to contain between 190 and 2145 astrophysical neutrinos. The
8-year diffuse muon neutrino flux results can be summarized as
\begin{gather*}
  \hat{\Phi} = 1.01 \times 10^{-18} \ \mathrm{GeV}^{-1}\mathrm{cm}^{-2}\mathrm{s}^{-1}\mathrm{sr}^{-1} \\
  \sigma_\Phi = 0.25 \times 10^{-18} \ \mathrm{GeV}^{-1}\mathrm{cm}^{-2}\mathrm{s}^{-1}\mathrm{sr}^{-1} \\
  \hat{\gamma} = 2.19 \\
  \sigma_\gamma = 0.1,
\end{gather*}
using the notation introduced in Equation~\eqref{eqn:obs_spec}, where
we also defined $E_0 = 100$~TeV. The point source analysis presented
in~\citet{Collaboration:2019db} is optimized to search for sources
with similar spectra to the observed diffuse astrophysical flux, and
the discovery potential for the Northern sky is on the level of
$\phi_\mathrm{dp} \sim
10^{-19}$~$\mathrm{GeV}^{-1} \ \mathrm{cm}^{-2} \ \mathrm{s}^{-1}$ for
a flux normalization at $E_0 = 100$~TeV.

These two results for the diffuse spectrum and point source search are
based on the same neutrino sample and are therefore particularly
well-suited to joint analysis. To reflect this sample, we use
$\delta_\mathrm{min}=-5^\circ$ in our joint model. As explained in
Section~\ref{sec:pdet_def}, we consider a minimum energy of
$E_\mathrm{min}^\mathrm{PS} = 1$~TeV and
$E_\mathrm{max}^\mathrm{PS} = 100$~PeV in the derivation of the
detection probability from the IceCube point source analysis. When
interpreting the results and comparing to previous work in
Section~\ref{sec:results}, we instead use $E_\mathrm{min}^L = 10$~TeV
and $E_\mathrm{max}^L = 10$~PeV as the energy range within which the
neutrino luminosity is defined, as it is not yet confirmed that the
power-law astrophysical flux measured by IceCube extends to other
energies, and this choice facilitates comparison with previous work.

The IceCube Collaboration has recently presented an updated point
source search using ten years of muon track
data~\cite{Aartsen:2020fm}. This work extends the analysis to both
hemispheres, but the Northern sky remains the strongest in terms of
sensitivity due to the reduced background to atmospheric neutrinos. In
the Northern sky, the discovery potential of this analysis is
comparable to that used here for the case of an $E^{-2}$ spectrum but
reduced by $\sim 30$\% for sources with a softer $E^{-3}$ spectrum. No
point sources are found to have a local significance of $> 5\sigma$,
although signals of $\sim 4\sigma$ have emerged. Due to the comparable
discovery potential and lack of $> 5\sigma$ detection, our results are
still relevant in light of this recent work.

\section{Results}
\label{sec:results}

\begin{figure*}[p]
  \centering
  \includegraphics[width=0.95\columnwidth]{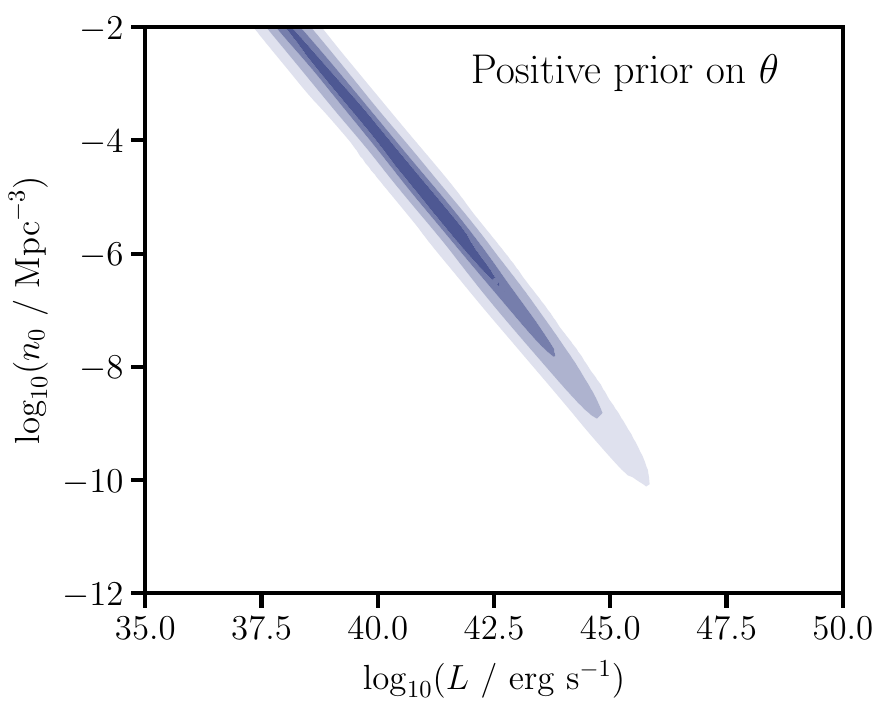}
  \includegraphics[width=0.95\columnwidth]{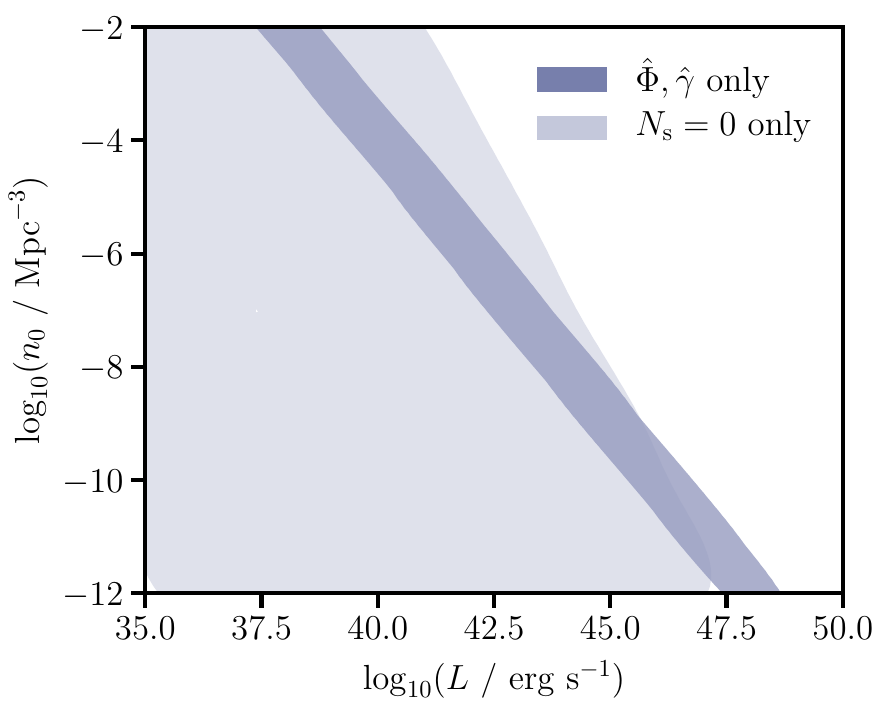}
  \includegraphics[width=0.95\columnwidth]{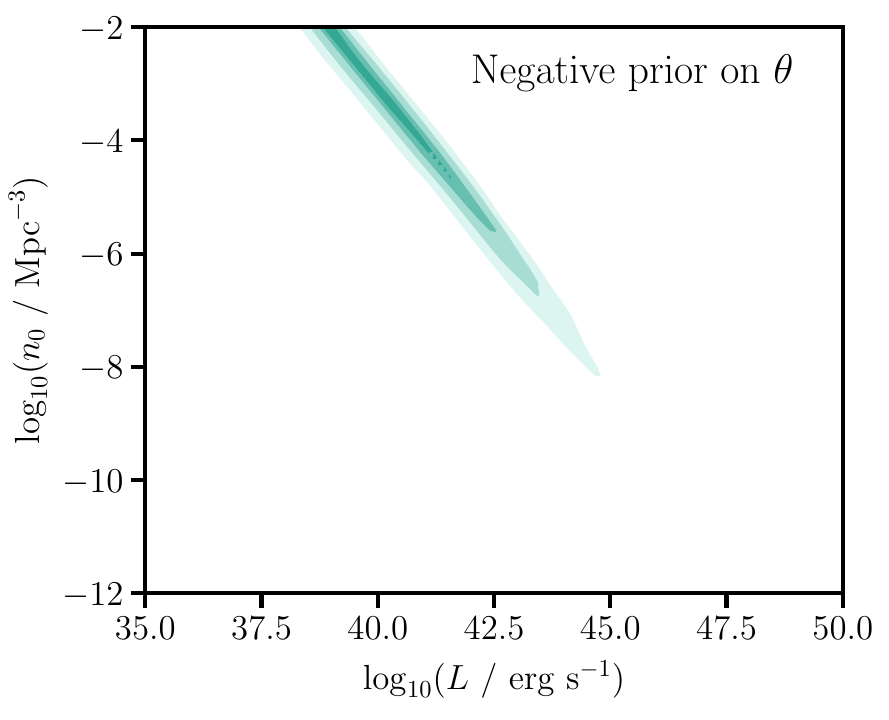}
  \includegraphics[width=0.95\columnwidth]{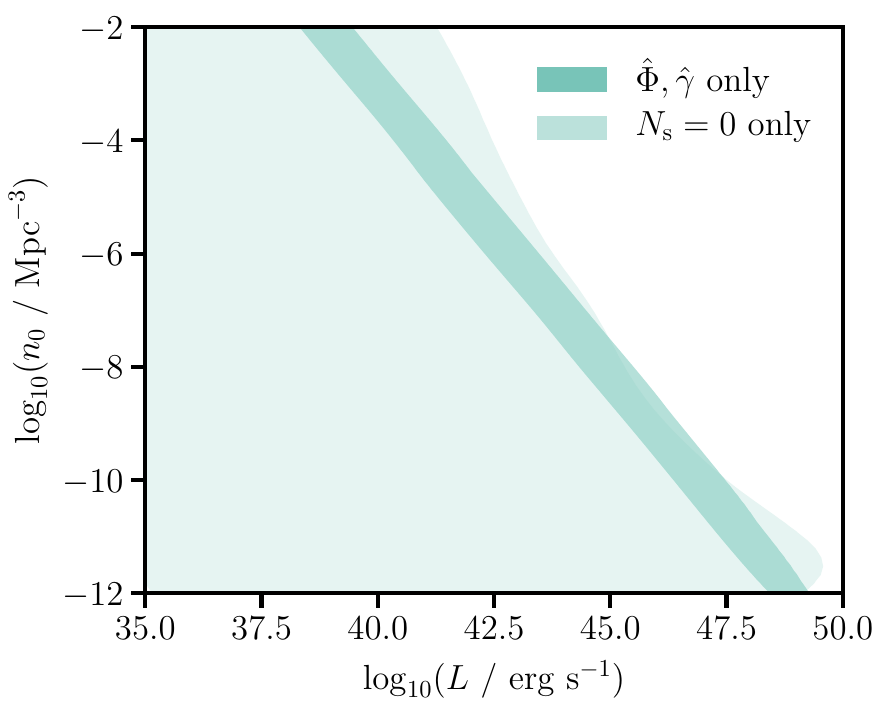}
  \includegraphics[width=0.95\columnwidth]{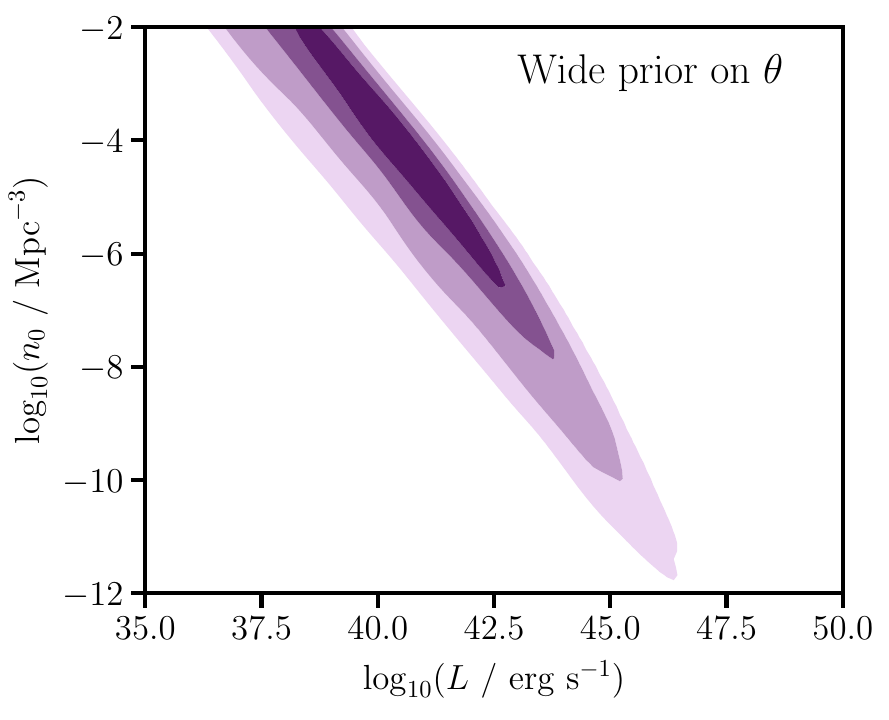}
  \includegraphics[width=0.95\columnwidth]{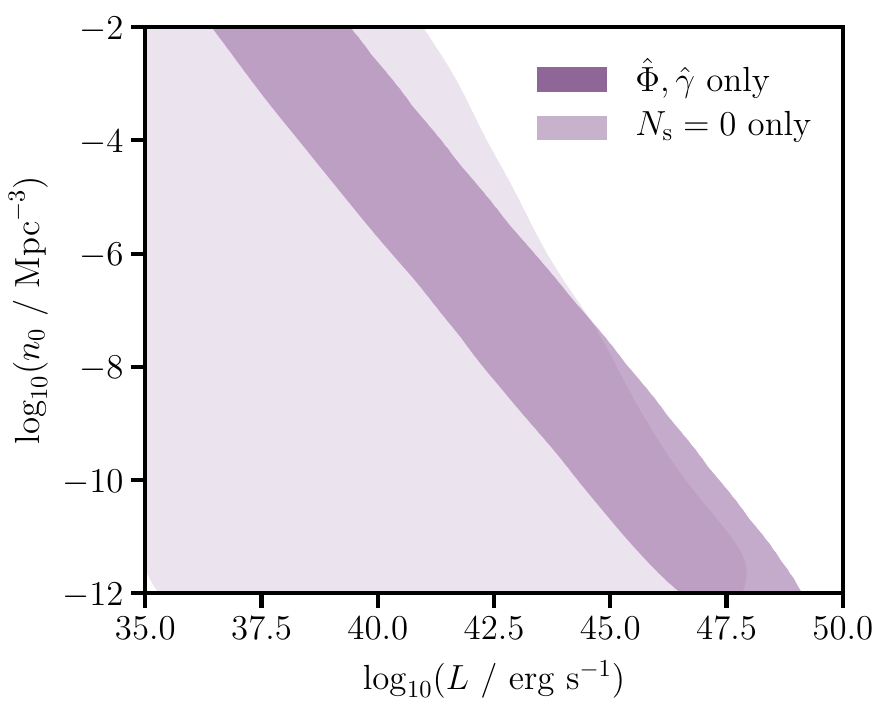}
  \caption{The joint posterior distribution for $L$ and $n_0$ under
    the observational constraints. The left panels show this
    distribution under the combined constraints for the case of a
    positive prior (top), a negative prior (middle) and a more
    unconstrained prior on $\theta$ (bottom), as detailed in
    Table~\ref{tab:priors}. The right panels give the same case as the
    left, with the two observational constraints plotted separately to
    show their relative contribution to the combined result. The
    constraint from requiring the source population to reproduce the
    observed astrophysical flux is labeled
    ``$\hat{\Phi}, \hat{\gamma}$ only'', and the constraint from the
    non-detection of significant point sources is given by
    ``$N_\mathrm{s} = 0$ only''. For the left plots, the contours show
    the 30, 60, 90, and 99\% highest posterior density credible
    regions for the joint distributions. In the right plots, only the
    99\% highest posterior density credible regions are shown. A
    kernel density estimate is used to plot the posterior samples and
    calculate the regions of highest posterior density. The
    non-observation of sources does not constrain low $L$ and large
    $n_0$ values, so the posterior continues unchanged in this
    regime.}
  \label{fig:joint_posterior}
\end{figure*}

The results can be summarized as the joint posterior distribution over
the model parameters. We show the joint and marginal posterior
distributions for the high-level population parameters $n_0$ and $L$
in Fig.~\ref{fig:joint_posterior}. In this way, the two-dimensional
joint distribution over $n_0$ and $L$ is analogous to the standard
presentations of the combined constraints from the observation of a
flux, but no point sources, that have been produced in several other
publications, including~\cite{Silvestri:2010ku, Murase:2012gy,
  Ahlers:2014em, Kowalski:2015gm, Murase:2016gly,
  Palladino:2019tm}. We see that the current observations disfavor a
population of rare, luminous sources, in agreement with what has been
found by previous studies of this nature (e.g.~\cite{Murase:2016gly,
  Collaboration:2019db, Palladino:2019tm}). Assuming positive source
evolution constrains the population to a narrow region of the
parameter space, with $n_0 \gtrsim 10^{-10}$~$\mathrm{Mpc}^{-3}$ and
$L \lesssim 10^{46}$~erg~$\mathrm{s}^{-1}$. Flat and negatively
evolving populations are more strongly constrained to
$n_0 \gtrsim 10^{-8}$~$\mathrm{Mpc}^{-3}$ and
$L \lesssim 5 \times 10^{44}$~erg~$\mathrm{s}^{-1}$. Considering a
more comprehensive range of possible source evolution shapes weakens
the constraints, as expected. Denser and less luminous populations are
most probable, but the posterior distribution generally has a long
tail out to more sparse populations of brighter sources. For
illustrative purposes, we also show the constraints on $n_0$ and $L$
separately in terms of the flux requirement and non-detection of
significant point sources in the right panels of
Fig.~\ref{fig:joint_posterior}. We see that the joint result is more
constraining than the individual contributions. The plots with
separate constraints are intended to aid the understanding of the main
result and connect to the presentation of results in previous work. We
emphasize that in our hierarchical model, the two constraints are
coupled through conditional probability statements to the shared
high-level parameters, and cannot be treated as independent. Indeed,
the posterior distribution is not well-identified in the
``$N_\mathrm{s} = 0$~only'' case, and sampling from it can result in
divergent transitions.

\begin{figure}
  \centering
  \includegraphics[width=\columnwidth]{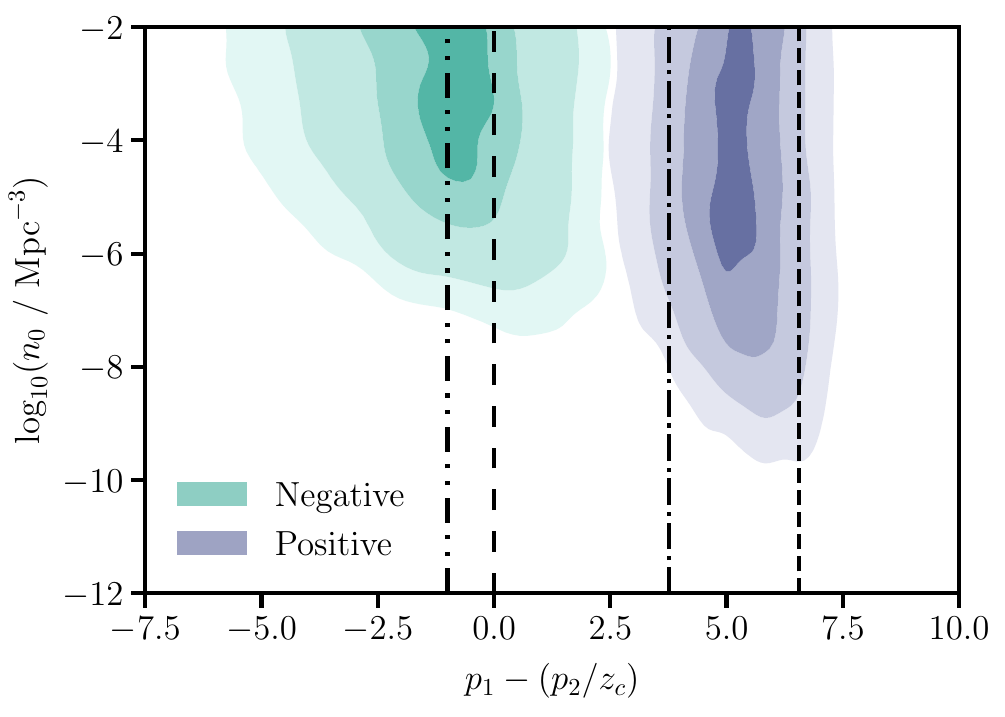}
  \includegraphics[width=\columnwidth]{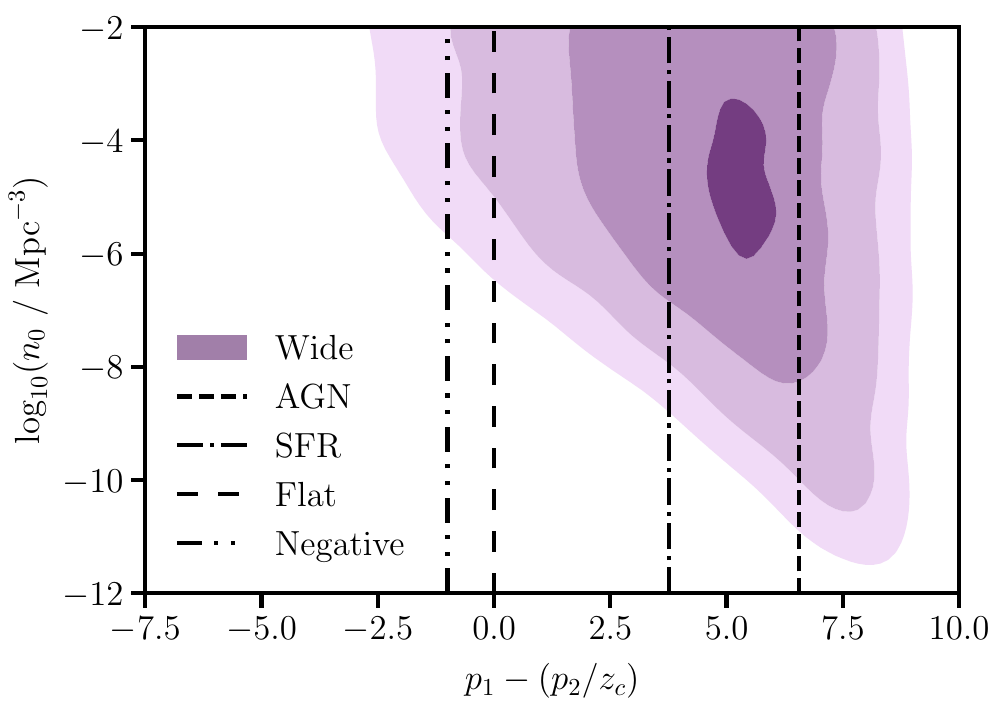}
  \caption{The dependence of $n_0$ on the source evolution for the
    three different evolution priors. We show $p_1 - (p_2/z_c)$ on the
    x-axis to summarize the dependence of $f(z, \theta)$ from
    Equation~\eqref{eqn:source_evolution} on these parameters at low
    redshifts ($z \ll 1$). The top panel shows the case of positive
    and negative evolution models and the bottom panel shows the case
    for the wide, unconstrained prior. In both panels we also plot the
    $p_1 - (p_2/z_c)$ values that correspond to commonly studied
    evolution models, as introduced in
    Fig.~\ref{fig:source_evolution}. The contours show the 30, 60, 90
    and 99\% highest posterior density credible regions, as in the
    left panels of Fig.~\ref{fig:joint_posterior}. It is clear that
    for more strongly evolving populations (higher values on the
    x-axis), $n_0$ is less constrained.}
  \label{fig:n0_evolution}
\end{figure}

We also show our results in terms of the dependency of the allowed
$n_0$ values on the source evolution in
Fig.~\ref{fig:n0_evolution}. Here, we show the same results as in the
left panels of Fig.~\ref{fig:joint_posterior}, but instead give the
joint posterior of $n_0$ and $p_1 - (p_2/z_c)$. $p_1$, $p_2$ and $z_c$
are parameters used to describe the redshift evolution of sources, as
introduced in Equation~\eqref{eqn:source_evolution}. The choice of
$p_1 - (p_2/z_c)$ reflects the behaviour of the evolution at low
redshift ($z \ll 1$) and therefore allows us to summarize the
$n_0$-evolution dependence in a simple 2D plot. We see that the
constraints for $n_0$ are very much dependent on the evolution model
and therefore the different priors chosen. Stronger evolution allows
for rarer source populations because in this case the contribution to
the observed neutrino flux from nearby sources is relatively small.

\begin{figure}[t]
  \centering
  \includegraphics[width=\columnwidth]{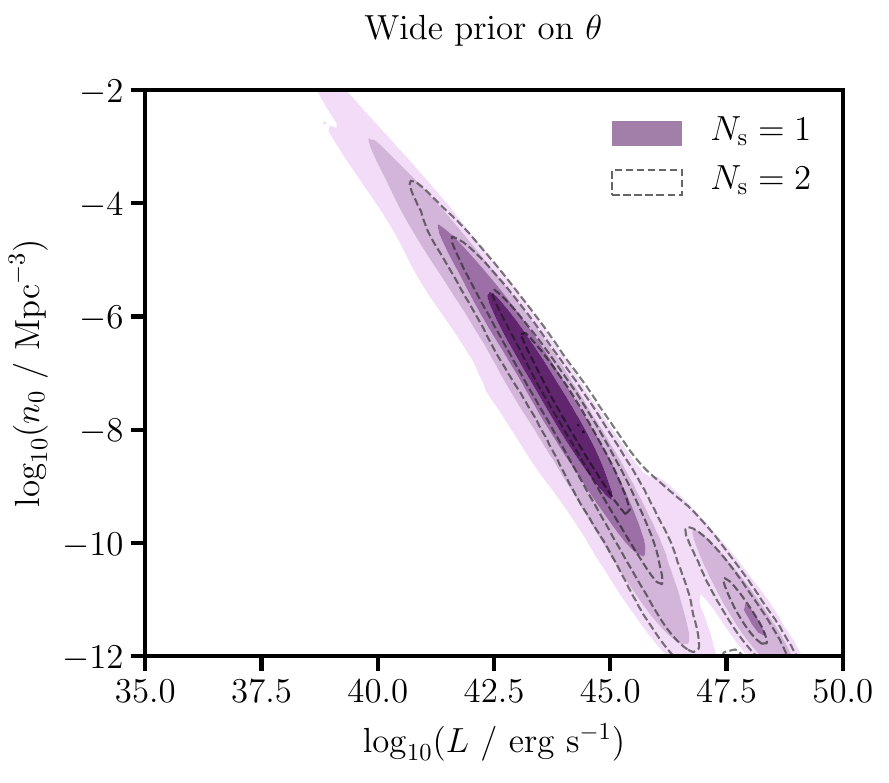}
  \includegraphics[width=\columnwidth]{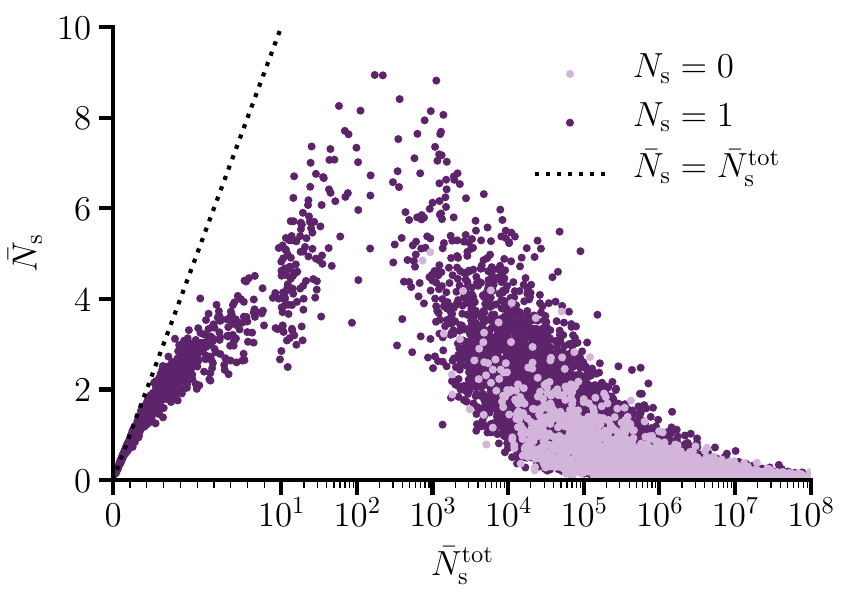}
  \caption{The case of an observation of one or two point sources with
    a pre-trial significance of above 5$\sigma$, with the current
    sensitivity and observed astrophysical flux. We consider an
    unconstrained prior on $\theta$ in the calculation of this
    result. The joint posterior distribution for $L$ and $n_0$ is
    shown in the upper panel, as in
    Fig.~\ref{fig:joint_posterior}. The lower panel shows samples from
    the joint distribution of the latent parameters
    $\bar{N}_\mathrm{s}^\mathrm{tot}$ and $\bar{N}_\mathrm{s}$ for the
    case of $N_\mathrm{s} = 1$ and $N_\mathrm{s} = 0$ for
    comparison. We see that the second peak in the upper panel at low
    $n_0$ and consequently low $\bar{N}_\mathrm{s}^\mathrm{tot}$
    corresponds to the possibility of a few bright sources being
    responsible for all the observed flux. This model is ruled out by
    the observation of near-isotropic emission, which is not included
    in our likelihood, as explained in the text.}
  \label{fig:joint_posterior_Ns_12}
\end{figure}

Fig.~\ref{fig:joint_posterior} shows the current source population
constraints, assuming the IceCube observation of a flux,
non-observation of point sources and the discovery potential described
in Section~\ref{sec:application}. To understand how these constraints
would be affected by the possible detection of point sources with the
current detector configuration and analysis methods, we also show the
results for the counter-factual case of $N_\mathrm{s} = 1$ and
$N_\mathrm{s} = 2$ in the upper panel of
Fig.~\ref{fig:joint_posterior_Ns_12}. For this hypothetical example,
we consider a population with an unconstrained evolution. The
detection of a single point source above the threshold would further
constrain the source population to values around
$n_0\sim10^{-8}$~$\mathrm{Mpc}^{-3}$ and
$L\sim10^{44}$~erg~$\mathrm{s}^{-1}$, with the population shifting to
less dense and brighter sources for further point source detections,
as can be seen from the shape of the posterior.

The bimodality in the joint posterior distribution reflects that it is
most probable for detected sources to come from a dense population,
but there is some non-negligible probability that a few rare, but very
bright sources could be responsible for all the observed flux under
our assumptions. We illustrate this point further in the lower panel
of Fig.~\ref{fig:joint_posterior_Ns_12} by showing the posterior
samples of the latent parameters $\bar{N}_\mathrm{s}^\mathrm{tot}$ and
$\bar{N}_\mathrm{s}$ for the case of $N_\mathrm{s} = 1$ (with
$N_\mathrm{s} = 0$ also shown for comparison). Large values of
$\bar{N}_\mathrm{s}^\mathrm{tot}$ directly corresponds to large $n_0$,
as shown in Equation~\eqref{eqn:total_sources}. Therefore, the samples
on the right side of this plot reflect the main peak in the upper
panel of Fig.~\ref{fig:joint_posterior_Ns_12}. As the total number of
sources decreases, the energy density implied by the observed flux is
shared among fewer sources. As such, each source is brighter, and more
sources are detectable. As $\bar{N}_\mathrm{s}^\mathrm{tot}$ continues
to decrease, there are so few sources in the universe that despite
their brightness, few are within the $P_\mathrm{det}$ horizon, and
eventually we reach the regime where
$\bar{N}_\mathrm{s} = \bar{N}_\mathrm{s}^\mathrm{tot}$, as marked by
the dashed line in the figure. However, this configuration of few
total sources is allowed in our model since by considering
$N_\mathrm{s} > 0$, we no longer require the neutrino arrival
direction to be close to isotropic. In constrast, for the case of
$N_\mathrm{s} = 0$, we see that only the high-$n_0$ component is
present, as it would be impossible to have a few bright sources with
no source detection. The bimodality arises from requiring that we
observe $N_\mathrm{s} = 1$, which imposes a strong constraint on high
$\bar{N}_\mathrm{s}$, as shown by the lower density of samples.  As
this requirement is relaxed and we consider $N_\mathrm{s} = 2$ and
higher, the two peaks move together and eventually merge, as indicated
in the upper panel of Fig.~\ref{fig:joint_posterior_Ns_12}.

While the second peak around $n_0 \sim 10^{-11}$~$\mathrm{Mpc}^{-3}$
and $L \sim 10^{48}$~erg~$\mathrm{s}^{-1}$ is permitted by these data,
it corresponds to the case of $\bar{N}_\mathrm{s}^\mathrm{tot} < 100$,
which is not physical for typical source classes considered. Even if
one point source were soon detected in IceCube, the overall
distribution of astrophysical neutrinos would still be close to
isotropic. These anisotropic models are still allowed in our analysis
because the observation of an isotropic flux is not explicitly
included into the data that we use. To include this information
mathematically, it would be necessary to extend the framework
described here to incorporate the individual neutrino event arrival
directions and energies. We also note that the total observed
astrophysical flux is
$\sim 10^{-17}$~GeV~$\mathrm{cm}^{-2}$~$\mathrm{s}^{-1}$ and the
discovery potential used in this work is around
$10^{-19}$~GeV~$\mathrm{cm}^{-2}$~$\mathrm{s}^{-1}$. So, even if all
source fluxes are at the discovery potential, we require at least
$4 \pi \Phi / \phi_\mathrm{dp} \sim 100$ sources to maintain the
observed flux and a near-isotropic distribution of detected neutrinos.

\begin{figure}[t]
  \centering
  \includegraphics[width=\columnwidth]{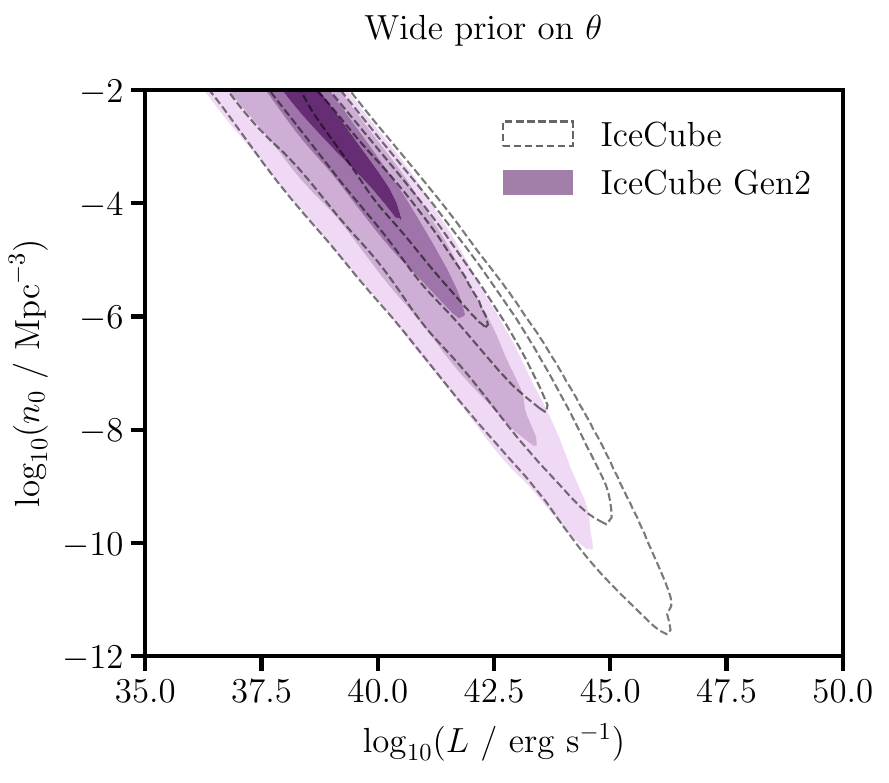}
  \caption{The joint posterior distribution for $L$ and $n_0$ for the
    case of IceCube Gen2, assuming no point source detections. The
    results are compared to the IceCube results, considering the case
    of an unconstrained prior, as shown in the bottom-left panel of
    Fig.~\ref{fig:joint_posterior}. The distributions are plotted as
    in Fig.~\ref{fig:joint_posterior_Ns_12}.}
  \label{fig:joint_posterior_gen2}
\end{figure}

Future detectors will improve the discovery potential of point source
analyses. The IceCube Collaboration plans IceCube Gen2, an expansion
of the instrumented volume of IceCube from 1~$\mathrm{km}^2$ to
10~$\mathrm{km}^2$ in the Antarctic ice \cite{Aartsen:2014njl}. The
planned design benefits from the increased understanding of the ice
properties from the development of IceCube, which makes it possible to
dramatically increase the spacing of the strings of photomultiplier
detector modules while maintaining the desired performance. By
increasing the exposure and angular resolution of the detector (thanks
to the longer muon track lengths that can be sampled in a larger
volume), there is an expected increase in the discovery potential of a
least $\sim 5$ compared to that of the current IceCube results for
15~years of Gen2 operation following 15 years of
IceCube~\citep{vanSanten:2017wy}. We rescale the detection probability
calculated in Section~\ref{sec:pdet_def} by a factor of 5 across all
declinations and show the equivalent source population constraints in
Fig.~\ref{fig:joint_posterior_gen2} for the case of the unconstrained
prior on the source evolution parameters. We see that for a continued
non-detection of significant individual sources, the constraints
become increasingly confined to numerous, dimmer sources, which would
produce an unresolvable diffuse flux in the detector. However, due to
the uncertainties present, there is still a long tail in the joint
distribution out to $n_0 \sim 10^{-10}$~$\mathrm{Mpc}^{-3}$ and
$L \sim 5 \times 10^{44}$~erg~$\mathrm{s}^{-1}$. The constraints that
we find for $n_0$ are similar to those reported using an angular power
spectrum analysis in~\citet{Dekker_Ando_2018}.

The KM3NeT Collaboration is also developing a cubic-kilometer scale
neutrino detector in the Mediterranean
Sea~\cite{AdrianMartinez:2016bf}. The detector is still under
construction and has recently started taking data successfully in
parallel with its continued expansion. With water as the detection
medium, the scattering length of photons in increased, and it is
possible to reconstruct the direction of such events with an angular
resolution of $<0.2^\circ$ at energies of $E>10$~TeV with a similar
module spacing to that of IceCube \cite{Aiello:2019kp}. This high
angular resolution means that the KM3Net detector will have an
improved discovery potential by a factor of $\sim3-4$ compared to
IceCube's Northern sky once it reaches its full size and equivalent
operation time (see Fig.~9 in \citet{Aiello:2019kp}). As the effective
area of KM3Net is very different from that of IceCube, in particular
in terms of declination dependence, applying our model to this case
would require a dedicated study of the detection probability, as
described in Section~\ref{sec:detection_probability}. We do not
attempt this here, but simply note that the potential impact on the
constraints would be somewhere between that of IceCube and IceCube
Gen2. The two detectors observe different regions of the sky, and so
their operation will be complementary in terms of point source
searches.

\subsection{Implications}

In order to understand the constraints for a particular class of
sources, we must model its luminosity distribution, as described in
Appendix~\ref{sec:lum_function}. Neutrino luminosity functions are
currently unconstrained for all candidate source classes, so it is
necessary to assume some connection between the measured
electromagnetic emission from a source class and its neutrino emission
in the context of a theoretical model. For this reason, we leave such
a comparison to future work and highlight the general implications of
our results here. A more in-depth discussion of relevant source
classes can be found in \cite{Ahlers:2014em} and
\cite{Murase:2016gly}.

\begin{figure}[t]
  \centering
  \includegraphics[width=\columnwidth]{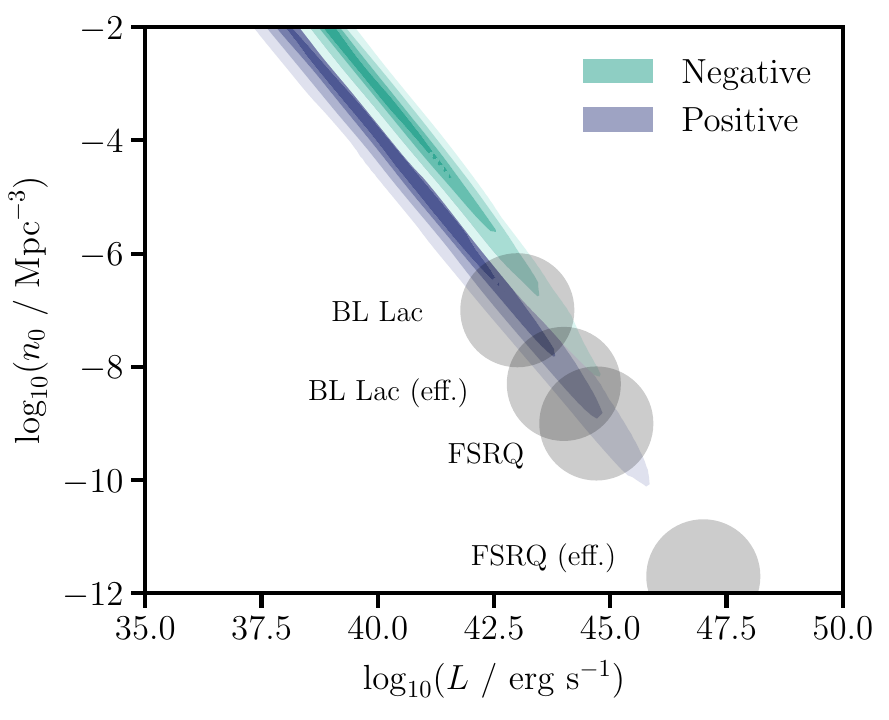}
  \caption{The joint posterior distributions are shown for the case of
    both positive and negative source evolution, as in
    Fig.~\ref{fig:joint_posterior}. We compare these distributions
    with expected values for BL Lacs and FSRQs that are shown by the
    shaded circles. These shaded areas are intended to guide the eye
    to values based on the results presented
    in~\citet{Ajello:2014fi}. The label ``eff.''\ denotes an estimate
    for the local density of these sources that is expected to
    dominate their contribution to the neutrino flux. This estimate is
    based on the effective luminosity and the formalism introduced
    in~\citet{Murase:2016gly}. The specific constraints for BL Lacs
    and FSRQs depend on the assumptions regarding the source
    evolution.}
  \label{fig:joint_posterior_implications}
\end{figure}

The general implications of our results are that populations of very
rare, luminous sources are disfavored as the primary contributors to
the observed astrophysical neutrinos, as otherwise, we would have
started to detect them as individual sources. This argument can be
weakened if we consider non-standard source evolutions, such as for
the case of the wide prior on $\theta$ in the bottom-left panel of
Fig.~\ref{fig:joint_posterior}, and particularly for the case of
rapidly evolving sources. For positively evolving source candidates
with a local density of $\gtrsim 10^{-6}$~$\mathrm{Mpc}^{-3}$, such as
starburst galaxies, AGN and galaxy clusters, the local density is
consistent with the 30\% region of highest posterior density. At lower
densities, possible candidates are energetic blazars such as BL Lac
objects. Their evolution is still not clear, with some studies
suggesting negative or flat cases~(see \citealt{Ajello:2014fi} and
references therein), which provide stronger constraints. Flat-spectrum
radio quasars (FSRQs) are popular source candidates with
$n_0 \lesssim 10^{-9}$~$\mathrm{Mpc}^{-3}$ that are widely agreed to
have a strong positive evolution. Following the assumptions on the
source evolution and luminosity function reported
in~\citet{Ajello:2014fi}, we compare the cases for these more
constrained sources to our results for positive and negative evolution
in Fig.~\ref{fig:joint_posterior_implications}. BL Lac objects lie in
the tail of our posterior distribution assuming negative evolution,
but are still consistent with our results if positively evolving
models are considered. FSRQs lie further out, and even including more
uncertainty into our model, it is difficult for such a population to
explain the IceCube observations without very strong evolution. These
results are in general agreement with those found from considering
more specific models for the individual source
classes~\citep{Murase:2016gly}, but do not disfavour BL Lacs and FSRQs
as strongly as previous work. It is important to remember that the
$n_0$ estimates for these sources depend on the evolution
reconstructed from their observations. In this way, further
investigation with more model-specific assumptions is necessary for us
to draw final conclusions on this matter, and we plan to explore this
direction in future work.

We find that the most probable region for the density of the neutrino
source population is $n_0 \geq 10^{-7}$~$\mathrm{Mpc}^{-3}$, which is
in general agreement with the results reported in previous
work. However, our results show that values down to
$n_0 \sim 10^{-11}$~$\mathrm{Mpc}^{-3}$ are consistent with the
current IceCube observations, depending on the source evolution. The
tail of the posterior distribution down to lower $n_0$ arises as we
model the detection probability gradient and include information on
uncertainty into our model, as we will discuss further below. Our
results are hence less constraining than those reported in Fig.~3 of
\citet{Murase:2016gly}, despite making use of the recent IceCube
observations and a more constraining detection criterion, as we
include more sources of uncertainty. In \citet{Murase:2016gly}, the
detection criterion is a multiplet above 50~TeV, including a
correction based on the estimated number of atmospheric neutrino
events, and the luminosity is defined as the differential energy flux
at 100~TeV. We note that there is no major difference in the
conclusions reached in our work if the luminosity is redefined in this
way. Fig.~4 of \citet{Murase:2016gly} also shows the case of using
harder blazar spectral templates, which is somewhat closer to our
results. As we mentioned in Section~\ref{sec:detection_probability},
there are now several multiplets above 50 and even 100~TeV, so the
criterion on which these results are based has now been surpassed. Our
results are also closer to, but still allow lower $n_0$ values than,
the SFR case shown in Fig.~4 of \citet{Palladino:2019tm}, which uses
multiplets above 200~TeV as a detection criterion, and Fig.~19 in
\citet{Collaboration:2019db}, which uses 90\% CL upper limits
calculated assuming $\gamma=2.19$.

In our work, we have explicitly modeled the detection probability of a
statistically significant signal in IceCube instead of assuming that
the observation of neutrino multiplets at high energies is a
sufficient requirement for detection. We utilize the additional
information in lower-energy neutrino events to uncover possible point
sources that may not yet result in a high-energy multiplet, making our
framework more sensitive. While there are higher backgrounds at lower
energies, the power-law nature of the expected source fluxes means
that this information is still useful in the statistical detection of
sources. Additionally, high-energy events can also appear as low
energy tracks that start outside the detector. If we ignore the
sources of uncertainty in our model and use the same assumptions
applied in previous studies, we actually find results that are more
constraining than those reported in~\citet{Collaboration:2019db}
and~\citet{Palladino:2019tm}. We cannot easily compare our approach
with that used in~\citet{Murase:2016gly} as the criterion for
multiplets above 50~TeV has now been exceeded in the publicly
available IceCube data. More generally, it is difficult to directly
compare the two methods as the IceCube point source analysis also
accounts for isotropic background fluctuations and the presence of
atmospheric neutrinos in the sample through a statistical
analysis. Given that we expect our definition of source detection to
result in stronger or comparable constraints to those shown in
previous results, our broader constraints can be understood in that we
have consistently accounted for relevant sources of uncertainty. The
current population of neutrino sources is unknown, as are important
population parameters such as the evolution and luminosity
function. Additionally, the astrophysical flux and spectral index
reconstructed from IceCube data also have associated uncertainties. By
modeling known sources of uncertainty, our results reflect a coherent
picture of our current state of knowledge with minimal modeling
assumptions.

The IceCube collaboration has recently reported an energetic neutrino
in coincidence with a blazar flare \cite{IceCube:2018dnn} and an
additional excess of lower-energy neutrinos from this point in the sky
\cite{Collaboration:2018kh}. The blazar, TXS~0506+056, was initially
identified as a BL Lac object but since been proposed for
reclassification as an FSRQ based on its radio and O\textsc{II}
luminosities, emission-line ratios and Eddington ratio
\cite{Padovani:2019vg}. While our results disfavor FSRQs as candidate
source classes for the majority of the astrophysical flux, we want to
emphasize that this is not in conflict with the first observation of
neutrino emission originating from these energetic sources. Our
results are relevant for the time-integrated emission from a
population of transient sources, whereas our assumptions do not
constrain single bright transients. Through coincident observation in
time, the instantaneous background to astrophysical neutrinos can be
drastically reduced, revealing sources that would be hidden in the
time-integrated emission.

\section{\label{sec:conclusions}Conclusions}

We have developed a Bayesian hierarchical model for the derivation of
constraints on a general population of neutrino sources. Within our
framework, we model the individual source detection probability and
propagate sources of uncertainty, allowing us to infer realistic and
model-independent constraints from the IceCube observations. For this
reason, our results are less constraining than those of previous
work. We present our results as a posterior distribution to facilitate
their interpretation and argue that rare sources with
$n_0 \gtrsim 10^{-10}$~$\mathrm{Mpc}^{-3}$ may be consistent with the
current IceCube observations for the case of strong source
evolution. However, even when including more sources of uncertainty,
it seems that rare, bright sources, such as BL Lacs and FSRQs remain
unlikely candidates. We also use our framework to show potential
constraints in the event of point source detection and the impact of
the expected improvement in the sensitivity of future detectors.

Our approach can be extended to include further relevant
information. Here, we have used muon track events in the Northern sky
as the most constraining channel, but this could be extended to
include other event types and lower declinations that can provide
complementary information thanks to their different properties.
Indeed, different spectral indices of the astrophysical component have
been found in different IceCube diffuse flux analyses (see e.g. Fig~13
in~\citet{Aartsen:2016hp}). The statistical formalism introduced here
could also be applied to constrain the rate and properties of
transient source populations.  Another compelling possibility is the
inclusion of multi-messenger observations. General constraints on the
power density of neutrino sources can be derived from the standard
phenomenological picture. For example, the Waxman Bahcall
bound~\cite{Waxman:1998dg} connects the average production rate of
ultra-high-energy cosmic rays to that of high energy neutrinos, and
the observations of the diffuse extra-Galactic gamma-ray background by
the Fermi-LAT instrument \cite{Ackermann:2015kc} constrain the
spectral index of the neutrino emission \cite{Murase:2013cq,
  Capanema:2020td}.  Moving away from a more model-independent view,
our framework can also be applied to a specific source scenario in
order to investigate the implications in detail. By defining a
theoretically motivated relationship between cosmic ray, neutrino, and
electromagnetic emission, the multi-messenger observations can be
combined coherently.

\section*{Acknowledgements}

We thank Kohta Murase for valuable comments on this manuscript and
Armando di Matteo for suggesting the presentation used in
Fig.~\ref{fig:n0_evolution}. We also acknowledge Hans Niederhausen,
Christian Haack, Lisa Schumacher and Martin Ha-Minh for useful
discussions related to IceCube, as well as Christer Fuglesang for
feedback on an early version of this manuscript. This work made use of
NumPy~\cite{numpy}, SciPy~\cite{scipy}, Astropy~\cite{astropy:2018},
Matplotlib~\cite{Hunter:2007ouj},
Seaborn~\cite{michael_waskom_2017_883859}, h5py
\cite{collette_python_hdf5_2014} and the High Performance Computing
facilities at the PDC center of the Swedish National Infrastructure
for Computing.

\appendix

\section{\label{sec:lum_function}Including a luminosity function}

The inclusion of a luminosity function is relatively straightforward,
once a suitable function is chosen. This function could either be
fixed to match observations, or parameterized to reflect reasonable
expectations. In the latter case, these parameters could be included
in the hierarchical model as has been done for the source
evolution. It would also be necessary to integrate over this function
in the calculations shown in Section~\ref{sec:physical_model} and the
subsequently derived expressions in
Section~\ref{sec:statistical_model}. From extending the diffuse
differential flux shown in Equation~\eqref{eqn:phi_n}, we would have
\begin{equation}
  \frac{\mathrm{d}\bar{N}_\nu^\mathrm{tot}}{\mathop{\mathrm{d}E} \mathop{\mathrm{d}t} \mathop{\mathrm{d}A} \mathop{\mathrm{d}\Omega}}
  = \frac{1}{4\pi} \int \mathop{\mathrm{d}L} \int_0^\infty \mathop{\mathrm{d}z}
  \frac{\mathrm{d}\bar{N}_\mathrm{s}}{\mathop{\mathrm{d}L} \mathop{\mathrm{d}V}}
  \frac{\mathrm{d}V}{\mathrm{d}z}
  \frac{\mathrm{d}\bar{N}_\nu^\mathrm{src}}{\mathop{\mathrm{d}E} \mathop{\mathrm{d}t} \mathop{\mathrm{d}A}},
\end{equation}
where
$\mathrm{d}\bar{N}_\mathrm{s}/ \mathop{\mathrm{d}L}
\mathop{\mathrm{d}V}$ is the luminosity function, including its
evolution with redshift. In the same way, the expected number of
detected sources shown in Equation~\eqref{eqn:expected_sources} would
be
\begin{equation}
  \bar{N}_\mathrm{s} = \frac{1}{4\pi} \int_\Omega \mathop{\mathrm{d}\Omega} \int \mathop{\mathrm{d}L} \int_0^{\infty} \mathop{\mathrm{d}z} P(\mathrm{det} | \phi, \gamma, \delta) \frac{\mathrm{d}\bar{N}_\mathrm{s}}{\mathop{\mathrm{d}L} \mathop{\mathrm{d}V}}\frac{\mathrm{d}V}{\mathrm{d}z}.
\end{equation}
If we substitute a luminosity function that is a delta function into
the above expression, we recover Equations~\eqref{eqn:phi_n} and
\eqref{eqn:expected_sources}. The effect of including the luminosity
function into the hierarchical model would be to increase the
constraints, with broader luminosity functions becoming more
constraining than the results presented above. This approach would be
well suited to assessing the viability of specific source classes
where it is possible to motivate a robust connection between the
observed electromagnetic radiation and the emitted neutrino flux.

\section{\label{sec:Pdet_Braun}Verification of the detection
  probability}

The specific implementation of the detection probability used in this
work is detailed in Section~\ref{sec:application}. Here, we illustrate
the general procedure used, following the approach in
\citet{Braun:2008kr} and verifying that we can reproduce the reported
results. We calculate the individual source detection probability for
the example case of a source at $\delta = 48^\circ$ with a power-law
spectrum. We use a simulated dataset of 67,000 background events from
an atmospheric-only isotropic component simulated down to
$E_\mathrm{min}^\mathrm{PS} = 100$~GeV. We sample the reconstructed
energies by using the reported marginal likelihood in Fig.~4 of
\citet{Braun:2008kr} and use the angular resolution
of~\citet{PSdataset:2018ab} with an offset of $0.2^\circ$ to reflect
the originally more pessimistic angular resolution assumptions.

\begin{figure}
  \centering
  \includegraphics[width=0.9\columnwidth]{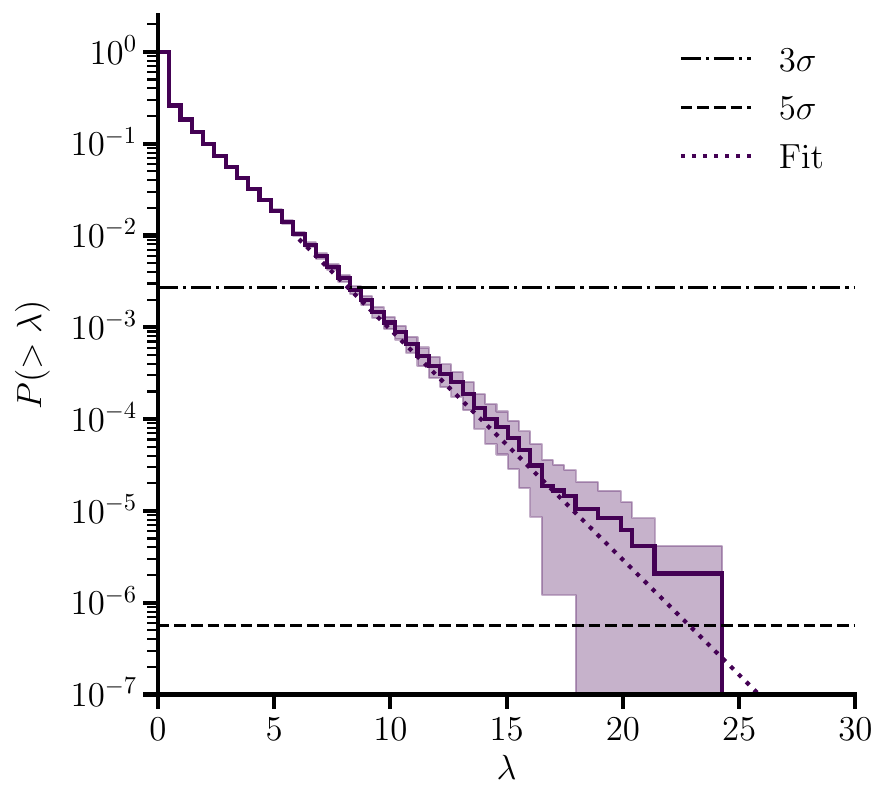}
  \caption{The $\lambda$ distribution for the simulated background is
    shown with the $\pm \sqrt{N_\mathrm{bin}}$ uncertainty as the
    shaded area, with $N_\mathrm{bin}$ the number of counts in each
    histogram bin (c.f. Fig.~6 in \citet{Braun:2008kr}). For this
    example we have calculated $\lambda$ in $4.8 \times 10^5$
    trials. The $3\sigma$ and $5\sigma$ levels are also plotted as the
    dash-dotted and dashed lines, respectively. An exponential
    function is fit to the tail of the distribution to determine
    $\lambda_{5\sigma}=22.8$, and this is also shown by the purple
    dashed line.}
  \label{fig:bg_TS}
\end{figure}

\begin{figure}
  \centering
  \includegraphics[width=0.9\columnwidth]{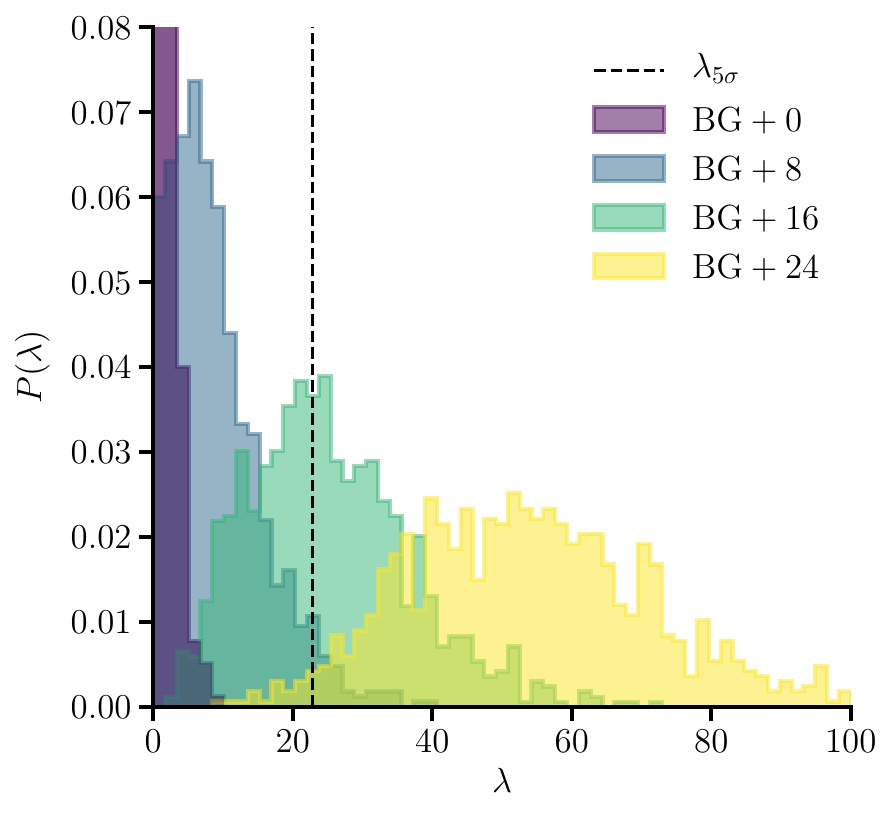}
  \caption{The $\lambda$ distribution for simulated background with a
    varying number of source events (c.f. Fig~6 in
    \citet{Braun:2008kr}). $10^3$ trials are shown for each case, and
    the injected signal has a power-law spectrum with $\gamma=2$. The
    dashed line shows $\lambda_{5\sigma}$, the threshold for a
    $5\sigma$ detection, and we see that injecting 16 source events
    corresponds to $P_\mathrm{det}\sim0.5$.}
  \label{fig:source_TS}
\end{figure}

\begin{figure}
  \centering \vspace{10pt}
  \includegraphics[width=0.9\columnwidth]{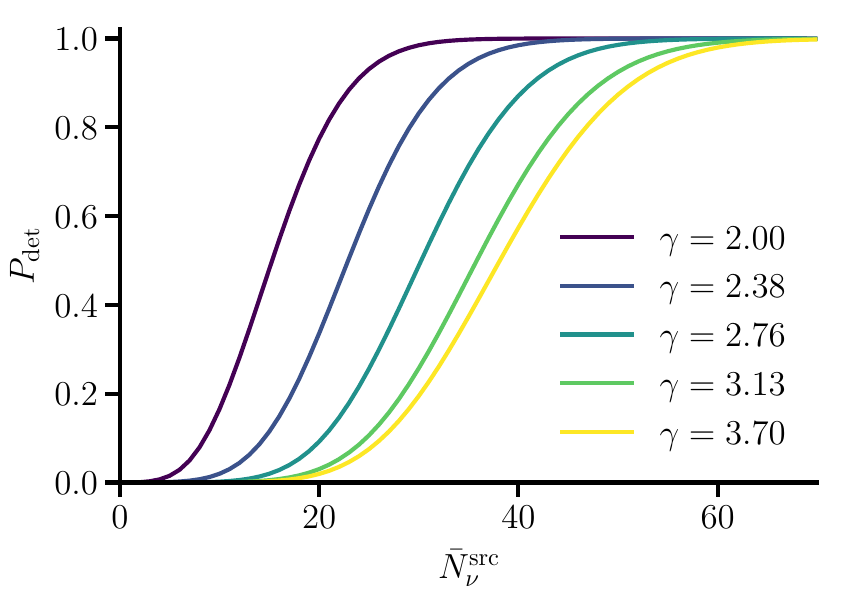}
  \caption{The detection probability as a function of the expected
    source counts, $\bar{N}_\nu^\mathrm{src}$. The case for $\gamma=2$
    is shown, based on the results presented in
    Fig.~\ref{fig:source_TS}, with steeper $\gamma$ shown for
    comparison. As expected, a steeper source spectrum requires more
    source counts for detection as it is more difficult to distinguish
    the signal events from the background (c.f. Fig.~7 and Fig.~8 in
    \citet{Braun:2008kr}).}
  \label{fig:Pdet_demo}
\end{figure}

First, we find $\lambda_{5\sigma}$ by evaluating the $\lambda$
distribution for a large number of background simulations. The
resulting cumulative probability distribution is shown in
Fig.~\ref{fig:bg_TS}. We then inject a fixed number of signal events
into the simulation and evaluate the $\lambda$ distribution in this
case, as shown in Fig.~\ref{fig:source_TS}. The detection probability
for a given mean number of source events, $\bar{N}_\nu^\mathrm{src}$,
is calculated by summing over the detection probabilities for source
events, $N_\nu^\mathrm{src}$, in the range 0 - 100 and weighting by
the Poisson probability,
$P(N_\nu^\mathrm{src} | \bar{N}_\nu^\mathrm{src})$. The resulting
detection probability is shown for a range of spectral indices in
Fig.~\ref{fig:Pdet_demo}.

For a given detector effective area and $\gamma$,
$\bar{N}_\nu^\mathrm{src}$ corresponds to the point source flux
normalization, $\phi$, as shown in
Equation~\eqref{eqn:aeff_folding}. Using the expression for the point
source flux given in Equation~\eqref{eqn:point_source_at_Earth}, we
see that
\begin{equation}
  P(\mathrm{det} | \bar{N}_\nu^\mathrm{src},\gamma,\omega_s) \Leftrightarrow P(\mathrm{det} | \phi,\gamma, \omega_s).
\end{equation}
In this way, the detection probability can be connected to the source
parameters described in Section~\ref{sec:physical_model}, and form
part of the generative model.

\section{\label{sec:atmosonly_Pdet}Impact of including $\mathbf{n_0}$
  and {\boldmath$\theta$} into the detection probability}

Changing the population density and evolution parameters, $n_0$ and
$\theta$, results in a variation of the diffuse astrophysical flux,
which is a background to point source detection. As we require the
diffuse astrophysical flux to match observations in our joint model,
in practice, the value of the diffuse astrophysical flux is somewhat
constrained, and this change will have a limited impact on our
results. However, in the low $n_0$ regime, we have the extreme case of
only a few sources in the population and effectively no diffuse
astrophysical background to source detection, leading to more
constraining results. Here, we explore this extreme case and
demonstrate that it has little impact on our final results.

\begin{figure}
  \centering
  \includegraphics[width=0.9\columnwidth]{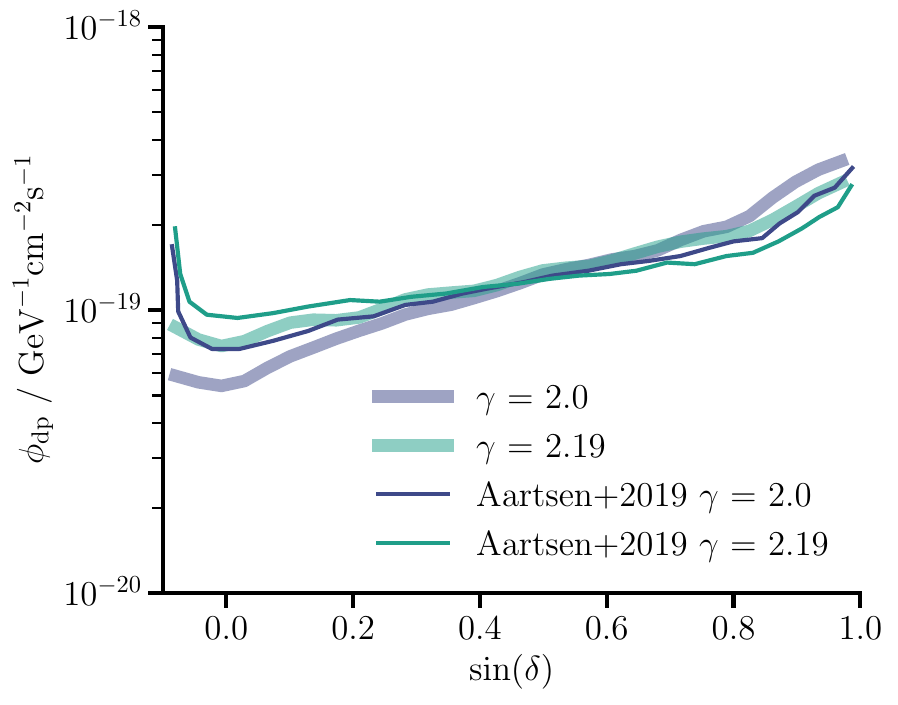}
  \caption{The discovery potential, as calculated in
    Fig.~\ref{fig:recreate_Aartsen+2019}, but for the case of a purely
    atmospheric diffuse background flux. }
  \label{fig:recreate_Aartsen+2019_atmosonly}
\end{figure}

Fig.~\ref{fig:recreate_Aartsen+2019_atmosonly} shows the discovery
potential in the case of a diffuse background that is purely due to
the atmospheric neutrino flux. The results are similar to what we find
in Fig.~\ref{fig:recreate_Aartsen+2019}, but we see a slight decrease
in $\phi_\mathrm{dp}$ at low declinations. This decrease is due to the
fact that at lower declinations, even the highest energy neutrinos are
not attenuated due to Earth-absorption, and by removing the
astrophysical background, we become more sensitive to high energy
point source signals.

\begin{figure}
  \centering
  \includegraphics[width=0.9\columnwidth]{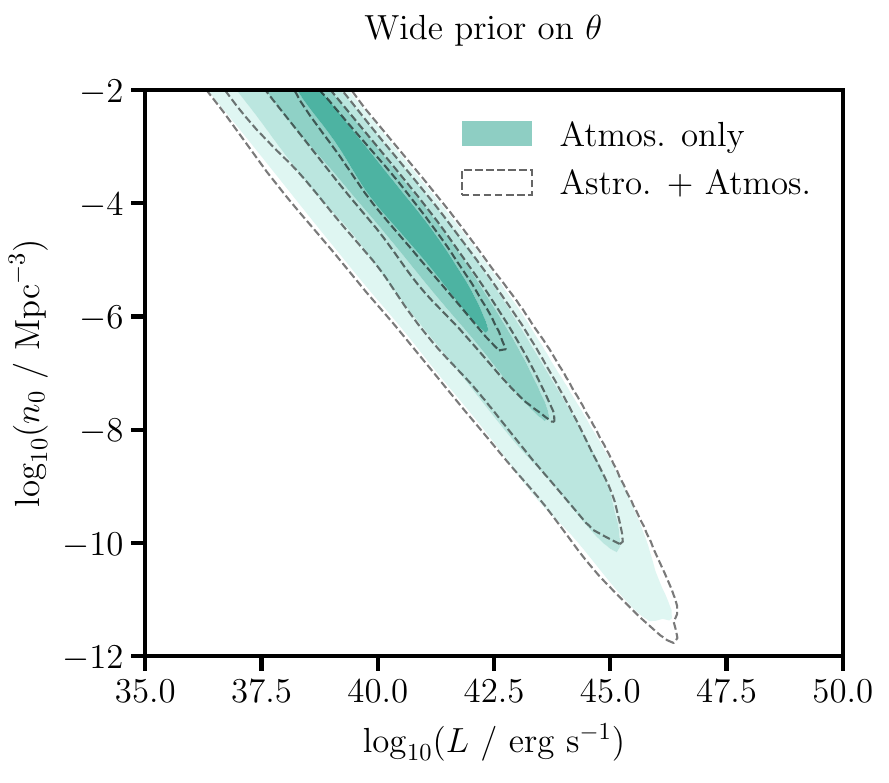}
  \caption{The joint posterior distribution, as in the bottom-left
    panel of Fig.~\ref{fig:joint_posterior}, but for the case of a
    purely atmospheric diffuse background flux in the detection
    probability calculation. The standard case of both astrophysical
    and atmospheric diffuse background components is also shown for
    comparison.}
  \label{fig:joint_posterior_atmosonly}
\end{figure}

We now calculate the detection probability for this case, as in
Section~\ref{sec:pdet_def}, and propagate this through into our final
results for the population parameters $n_0$ and $L$, which are shown
in Fig.~\ref{fig:joint_posterior_atmosonly}. We see good agreement
between the two cases with no apparent differences.

\FloatBarrier

\begin{widetext}
  \section{\label{sec:full_posterior_derivation}Full posterior
    derivation}

  In Section~\ref{sec:implementation}, we present the expression in
  Equation~\eqref{eqn:posterior} as the form of the posterior
  distribution. Here, we derive this expression in full. The posterior
  distribution is defined over the model parameters, $\theta$, $n_0$,
  $L$, $\bar{N}_\mathrm{s}$ and $\bar{N}_\mathrm{s}^\mathrm{tot}$, and
  is conditional on the observations, $N_\mathrm{s}$, $\hat{\Phi}$,
  $\hat{\gamma}$, and the model assumptions, $M$. This gives us
  \begin{equation}
    P(\theta, n_0, L, \gamma, \bar{N}_\mathrm{s}, \bar{N}_\mathrm{s}^\mathrm{tot} \Phi | N_\mathrm{s}, \hat{\Phi}, \hat{\gamma}, M) \propto P(\theta, n_0, L, \gamma, \bar{N}_\mathrm{s}, \bar{N}_\mathrm{s}^\mathrm{tot}, \Phi | M) P(N_\mathrm{s}, \hat{\Phi}, \hat{\gamma}| \theta, n_0, L, \gamma, \bar{N}_\mathrm{s}, \bar{N}_\mathrm{s}^\mathrm{tot}, \Phi, M),
  \end{equation}
  where we have used Bayes' theorem to expand the posterior. In a
  complex hierarchical model with many latent parameters, it is not
  always clear how to factorize terms into the prior and likelihood
  function, as typically done for simpler examples. However, for the
  purpose of this derivation, we consider the first term to be the
  joint prior distribution, and the second term the joint
  likelihood. We can proceed by expanding the prior term using the
  chain rule, to separate out the hyperparameters from the latent
  parameters. This gives
  \begin{equation}
    P(\theta, n_0, L, \gamma, \bar{N}_\mathrm{s}, \bar{N}_\mathrm{s}^\mathrm{tot}, \Phi | M) = P(\theta, n_0, L, \gamma | M) P(\bar{N}_\mathrm{s}, \bar{N}_\mathrm{s}^\mathrm{tot}, \Phi | \theta, n_0, L, \gamma),
  \end{equation}
  where the first term is the joint prior over the hyperparameters. In
  our model, we treat $\theta$, $n_0$, $L$ and $\gamma$ as
  independent, and this term factorizes into independent
  probabilities. In expanding the second term , we have also used
  conditional independence to write this as independent of $M$. The
  second term can also be factorized further using the conditional
  independence of $\bar{N}_\mathrm{s}$,
  $\bar{N}_\mathrm{s}^\mathrm{tot}$ and $\Phi$ such that
  \begin{equation}
    P(\theta, n_0, L, \gamma, \bar{N}_\mathrm{s}, \bar{N}_\mathrm{s}^\mathrm{tot}, \Phi | M) = P(\theta, n_0, L, \gamma | M) P(\bar{N}_\mathrm{s} | \theta, n_0, L, \gamma) P(\bar{N}_\mathrm{s}^\mathrm{tot} | n_0, \theta) P(\Phi | \theta, n_0, L, \gamma).
  \end{equation}
  We continue by now considering the likelihood factor in our original
  expression for the posterior distribution. Again, we start by
  expanding the expression using the chain rule and separating out
  independent observations from those that are connected, which yields
  \begin{equation}
    P(N_\mathrm{s}, \hat{\Phi}, \hat{\gamma}| \theta, n_0, L, \gamma, \bar{N}_\mathrm{s}, \bar{N}_\mathrm{s}^\mathrm{tot}, \Phi, M) = P(\hat{\gamma} | \theta, n_0, L, \gamma, \bar{N}_\mathrm{s}, \bar{N}_\mathrm{s}^\mathrm{tot}, \Phi, M) P(N_\mathrm{s}, \hat{\Phi} | \theta, n_0, L, \gamma, \bar{N}_\mathrm{s}, \bar{N}_\mathrm{s}^\mathrm{tot}, \Phi, M).  
  \end{equation}
  We now make use of conditional independence of $\hat{\gamma}$,
  $N_\mathrm{s}$ and $\hat{\Phi}$ to simplify this expression as much
  as possible
  \begin{equation}
    P(N_\mathrm{s}, \hat{\Phi}, \hat{\gamma}| \theta, n_0, L, \gamma, \bar{N}_\mathrm{s}, \bar{N}_\mathrm{s}^\mathrm{tot}, \Phi, M) = P(\hat{\gamma} | \gamma) P(N_\mathrm{s} | \bar{N}_\mathrm{s}) P(\hat{\Phi} | \bar{N}_\mathrm{s}^\mathrm{tot}, \Phi).  
  \end{equation}
  In the regime of large $n_0$, $\hat{\Phi}$ is independent of
  $N_\mathrm{s}^\mathrm{tot}$. However, in the low $n_0$ limit, we
  enter the regime where could be very few sources in the observable
  universe, and the detected astrophysical flux is conditionally
  dependent on the presence or lack of sources. We can treat this by
  expanding the final term as a mixture model over the case where
  $N_\mathrm{s}^\mathrm{tot} = 0$ and its complement, as shown in
  Equation~\eqref{eqn:mixture_model}. Bringing this all together, we
  have the complete expression shown in
  Equation~\eqref{eqn:posterior}.
  \begin{equation}
    \begin{split}
      P(\theta, n_0, \gamma, \bar{N}_\mathrm{s}, &\bar{N}_\mathrm{s}^\mathrm{tot}, \Phi | N_\mathrm{s}, \hat{\Phi}, \hat{\gamma}, M) \\
      & \propto P(\theta, n_0, L, \gamma | M) P(\bar{N}_\mathrm{s} |
      \theta, n_0, L, \gamma) P(\bar{N}_\mathrm{s}^\mathrm{tot} | n_0,
      \theta) P(\Phi | \theta, n_0, L, \gamma) P(\hat{\gamma} |
      \gamma) P(N_\mathrm{s} | \bar{N}_\mathrm{s}) P(\hat{\Phi} |
      \bar{N}_\mathrm{s}^\mathrm{tot}, \Phi).
    \end{split}
  \end{equation}

\end{widetext}

\bibliography{nu_pop_constraints}{}

\end{document}